\documentclass[twocolumn]{aastex631}
\usepackage{pgffor}
\usepackage{amsmath}
\hypersetup{allcolors=blue!20!black}

\sloppypar
\definecolor{g2}{rgb}{0.0, 0.5, 0.0}

\newcommand{\gaia}{\textsl{Gaia}}
\newcommand{\tmass}{\textsl{2MASS}}
\newcommand{\unwise}{\textsl{unWISE}}
\newcommand{\allwise}{\textsl{AllWISE}}
\newcommand{\wise}{\textsl{WISE}}
\newcommand{\ariadne}{\textsl{Ariadne}}

\shorttitle{a data-driven search for infrared excesses}
\shortauthors{contardo et al.}

\begin{document}

\title{\Large%
A Data-Driven Search For Mid-Infrared Excesses Among Five Million Main-Sequence FGK Stars}

\author[0000-0002-3011-4784]{Gabriella Contardo}
\email{gabriella.contardo@sissa.it}
\affiliation{Theoretical and Scientific Data Science, Scuola Internazionale Superiore di Studi Avanzati (SISSA), Trieste, Italy}

\author[0000-0003-2866-9403]{David W. Hogg}
\affiliation{Center for Cosmology and Particle Physics, Department of Physics, New~York~University, New~York NY, USA}
\affiliation{Center for Computational Astrophysics, Flatiron Institute, New~York NY, USA}
\affiliation{Max-Planck-Institut f\"ur Astronomie, Heidelberg, Germany}

\begin{abstract}\noindent
Stellar infrared excesses can indicate various phenomena of interest, from protoplanetary disks to debris disks, or (more speculatively) techno-signatures along the lines of Dyson spheres.
In this paper, we conduct a large search for ``extreme'' infrared excesses, designed as a data-driven contextual anomaly detection pipeline.
We focus our search on FGK stars close to the main sequence to favor non-young host stars. We look for excess in the mid-infrared, unlocking a large sample to search in while favoring extreme IR excess akin to the ones produced by Extreme Debris Disks (EDD) and/or planetary collision events.
We combine observations from ESA \gaia{} DR3, \tmass{}, and the \unwise~version of NASA \wise{}, and create a catalog of 4,898,812 stars with $G<16$\,mag. We consider a star to have an excess if it is substantially brighter in $W1$ and $W2$ bands than what is predicted from an ensemble of machine-learning models trained on the data, taking optical and near-infrared information as input features.
We apply a set of additional cuts (derived from the ML models and the objects' astronomical features) to avoid false positives and identify a set of 53 objects, including one previously identified EDD candidate.
Typical infrared-excess fractional luminosities we find are in the range $0.005$ to $0.1$, consistent with previous EDDs candidates and potential planetary collision events. 
\end{abstract}

\keywords{}


\section{Introduction}
\label{sec:intro}
The presence of infrared excess in stellar objects is, in general, a sign of circumstellar dust, protoplanetary disks, or debris disks. The strength of IR excesses is expected to decrease as a function of stellar age as a star's disk evolves \citep{wyatt2015evoldisks}, tying it to the evolutionary process in the star, its potential planets and other planetesimals or asteroid belts. As such, strong IR excesses are often associated with (relatively) young stellar objects. \cite{cotten2016census} detect IR excess in main-sequence stars, with a range of fractional luminosity $L_{IR}/L*$ between $10^{-5}$ to $10^{-2}$, averaging around $10^{-4}$, and with a single candidate having a fractional luminosity $ \simeq 0.1$. However, several studies have also highlighted the existence of high IR excess (with fractional luminosities around 0.02 and up to 0.1) for stellar objects with wider age ranges \citep{moorEDD, balog2009EDD, kennedy2013exozodi, song2005collisions, zuckerman2012circumstellar, melis2021structureddisktransit, rhee2008collision}, up to Gyrs. These fractional luminosities are in between the ones expected for (regular) debris disks ($L_{IR}/L*< 10^{-3}$) and protostars ($L_{IR}/L* > 0.1$) \citep{betapicturi2005uzpen}. Those excesses are currently thought to come from planetary or planetesimal collisions and often referred to as Extreme Debris Disks (EDDs). However, the stellar ages of some of these candidates are in tension with the current models of rocky planet formation which predict that most collision events (leading to such disks) should occur within the first 100 Myr. Therefore, these objects might indicate that these processes last longer than thought or that other processes lead to such disks. 

\cite{kenworthy2023planetscollision} recently identified a Sun-like object of 300 Myr undergoing a sharp increase in infrared, reaching a fractional luminosity similar to the ones stated above ($0.04$), followed by an obscuration in the optical some 1,000 days later. The authors show that the observations are consistent with a planetary collision between two exoplanets of several to tens of Earth masses at 2–16 astronomical units from the central star. Another Sun-like star, of $\sim$ 600Myr, with a detected IR-excess of $0.01$ fractional luminosity with an apparent dimming in the optical was identified in \cite{melis2021structureddisktransit}. Interestingly, there is also one account of a fast disappearance of the IR-excess \citep{melis2012disappearanceIRexcess}.



Currently, there are very few known EDD candidates: for instance, there were 17 reported in \cite{moorEDD}. Those events are a relatively rare occurrence: \cite{moorEDD} searched through $78,650$ objects and found 8 candidates --6 new and 2 known, i.e. an ``occurrence''\footnote{We note that due to the way some of the searches (including this one) are done --usually involving a variety of threshold-based or hard cuts, it is difficult to define properly the occurrence rate of those excesses.} rate of 0.01\%. A similar occurence rate was suggested by \cite{kennedy2013exozodi}. Those extreme excesses thus likely come from rare or potentially very short-lived events. It is critical to increase our sample of candidates to better understand the underlying physical causes of these extreme IR excesses. 

Interestingly, another possible---albeit less likely---causes of extreme IR excess are techno-signatures such as Dyson Spheres (hypothetical megastructures built around stars to collect energy). For instance, \cite{suazo2022hephaistos} provided upper limits on the prevalence of partial Dyson Spheres combining data from \gaia{} DR2 and \allwise{}, using models of the resulting IR-excess (and optical obscuration) of different types (stages and temperature) of Dyson Spheres. Following on this work, \cite{suazo2024} recently presented a search targeting such IR-excess types relying on these Dyson spheres models, with 7 candidates identified. The search for (extreme) contextual outliers in the IR is a nice example where the search for extraterrestrial intelligence (SETI) field intersects and can challenge the boundaries of current astrophysical models (in this instance, of planetary formation and evolution). As one would need to investigate and eliminate all other explanations before concluding an extreme IR-excess source is a Dyson Sphere, searches for ``Dysonian'' signatures will potentially provide interesting objects for ``mainstream'' astrophysics. This aspect has been coined as the ``ancillary benefit'', one of the axes of merits one can evaluate technosignature searches by, defined by \cite{sheikh2020axes}. 

We propose here to conduct a large search for such ``extreme'' infrared excesses.
We focus on a population of photometrically main-sequence stars, to favor potentially non-young or transitional candidates. We also restrict our search to FGK stars, focusing on Sun-like systems: as this work is partially motivated by the search for techno-signatures, as well as potential planetary collision of the like found in \cite{melis2010incidencerate, kenworthy2023planetscollision}, we want to focus on systems similar to ours. Additionally, reducing the variety of stellar types is also motivated by the nature of our approach. A larger diversity of stellar types would require our models to learn this and could lead to biases in our detection in terms of stellar types. However, the following methodology could and should be explored on other stellar populations.

We choose to conduct our search in the mid-IR, and more specifically we look for excess in the \wise{} $W1$ and $W2$ bands: It has been shown that IR-excess typical of EDDs can be hot enough to show in $W1$ and $W2$, and we expect (possibly extremely naively) to see techno-signatures near the water triple-point. Restricting to these two \wise{} bands also unlocks a larger sample to investigate, as the sensitivity and reliability of \wise{} is much higher in those bands than in the $W3$ and $W4$ bands: our curated dataset, presented in Section \ref{sec:data}, has 4.9 million objects. On the other hand, looking for excess in $W1$ and $W2$ means that we target only extreme events leaking in those bands, which can nonetheless be subtle, especially in the $W1$ band. 

To approach this problem in a way that scales to our catalog, we present a data-driven pipeline. Our approach alleviates the need for good (computationally expensive) stellar modeling or fitting: we are looking instead for deviations (or contextual anomalies) \emph{according to the data}, i.e., our sample of main-sequence FGK stars. That is, instead of using stellar modeling or approximations, we train a machine learning model to predict the expected $W1$ and $W2$ magnitudes from other (optical and near-infrared) photometric features. We then use this model to find stars with observed magnitudes in $W1$ and $W2$ higher than predicted (similar to what one would do with regular stellar modeling). Additionally, we deploy a set of measures to focus our search on stars that have consistently wrong predictions when using models trained on different subparts of the data. Details on our methodology are provided in Section \ref{sec:method}. 
This approach is also computationally interesting compared to ``proper'' SED fitting to scale to our 4.9 million objects dataset. 

From a methodological perspective, this paper presents a pipeline to conduct targeted, contextual anomaly detection (see \cite{chandola2009anomaly} for an anomaly-detection taxonomy overview), benefiting from robust supervised machine learning methods (here Random Forest). Anomaly detection (in astronomy and astrophysics) is often framed as finding rare objects, i.e., finding data examples with low probability (in low-density regions) according to the data. To tackle this problem with ML methods, one can (for instance) get a density estimate, and retrieve the objects with low scores. Some caveats of this generic approach are that (i)~obtaining reliable density estimates is non-trivial, especially as the number of dimensions increases, and (ii)~such an approach will retrieve all (potentially) rare/odd objects, but not necessarily the ``interesting ones'' (which motivates the growing literature on integrating active-learning in anomaly detection pipeline in astronomy such as in \cite{ishida2021active} and \cite{lochner2021astronomaly}). However, it is a good way to potentially uncover the infamous ``unknown unknowns''. On the other hand, in many cases, we might be looking for ``known unknowns'' (as in well-specified unknowns, like here, where we want to look for specific types of anomalies: stars with IR-excess). ``Knowing'' our unknowns, however, does not necessarily mean that we can turn this problem into a supervised classification one (where our detector would tell us a binary answer or a score probability of our object being an anomaly): We might not have actual examples of these anomalies, or only very few examples.

Here we propose to benefit from the fact that we are looking for targeted anomalies that deviate in a specific way (excess), in a specific region of the feature space (mid-infrared), considering their ``context'' (their optical and near-IR observations). We will be looking for objects that are confidently wrong in the prediction of that part of the feature space (the MIR-bands) using the rest of the feature space (mostly optical and near-IR photometry, see Section \ref{sec:method} for details). This process is indirectly doing a form of density estimation: as we assume our model is flexible enough and generalizes well, objects with a high, confident error, deviate from the training set distribution in some significant ways, that are not predictable from their ``context'' (input features). We note however that if the objects we are looking for were very common in our sample and with an excess predictable from the input features, our pipeline would not flag them as anomalies.

One amusing thing about taking a data-driven approach is that we require less, and less accurate, knowledge about how stars work.
Most infrared excess projects will have to deal with dust reddening and have physically accurate models for the multi-wavelength emission from stars.
We require neither; we require only that infrared excess events be around stars that are typical in the sense that there are (many) other stars in the target sample that are similar but lacking an infrared excess.

The rest of this paper is organized as follows: the data curation is described in Section \ref{sec:data}. The method details are provided in Section \ref{sec:method} along with a list of subsequent cuts applied to obtain our final set of 53 candidates, which include one previously detected EDD candidate (see Appendix \ref{sec:appendix:overlap}). Further analysis of these candidates including checking for H$\alpha$ emission, variability, Black Body fitting and their parameters, and rate of recovery for those, can be found in Section \ref{sec:analysis}. We provide concluding remarks in Section \ref{sec:discu}. We provide in the Appendix \ref{sec:appendix:canditable} a summary Table of our candidates, and in Appendix \ref{sec:appendix:deficit} a check for mid-IR deficit. The data products as well as the pipeline code and the analysis code can be found at \url{https://github.com/contardog/NotATechnosignatureSearch}.

\section{Data}
\label{sec:data}

We combine data from \gaia~DR3 \citep{gaia2016, gaia2023, gaia2023b}, the Two Micron All Sky Survey (\tmass) \citep{2006tmass} \citep{2006tmass},  the \unwise~catalogue \citep{2019unwise} and \allwise~catalogue \citep{2014AllWiseCatalogVizier}. We describe in this section our data curation and selection pipeline.

Using \gaia~DR3 observations, we select stars with a $G$-band magnitude brighter than $G < 16 mag$. We apply a temperature cut s.t. the effective temperature is between $4000$ and $7000$K (restricting the stellar types to FGK), using effective temperatures from \gaia~DR3 (\textit{teff\_gspphot}, inferred by GSP-Phot Aeneas from BP/RP spectra, apparent G magnitude and parallax). We impose the parallax over error to be over $10$, ruwe $< 1.4$  and require that \textit{ebpminrp\_gspphot} (reddening value using BP/RP spectra) is not null.
\
We rely on the cross-match between \gaia~DR3 and \tmass~(\textit{gaiadr3.tmass\_psc\_xsc\_best\_neighbour}), and between \gaia~DR3 and \allwise~(\textit{gaiadr3.allwise\_best\_neighbour}), provided by the ESA, to get the counter-part candidates in each catalog. Using the DPAC curated catalogs available on ESA \gaia~archive, we impose that the contamination flags (\textit{cc\_flags}) from \allwise~is '$0000$', i.e. there is no (flagged) contamination in any of the bands. We also restrict our sample to enforce that the photometric uncertainty in \tmass~is under $0.2$ for all bands, and with a \tmass~ photometric quality flag of 'AAA' (i.e. SNR $> 10$).
Additional cuts to prevent potential mismatch or contamination are also applied: we enforce that the \textit{``number of mates''} in both \tmass~and \allwise~is 0 (i.e. there is no other \gaia~sources that have this object as best neighbor as well: this should ensure (to some extent) that we do not get sources that were actually more than 1 object but unresolved in \tmass~or \wise~), and that the \textit{``number of neighbors''} is equal to 1 (i.e. there is no additional source in the external catalog that matches the \gaia~source within position errors). Both columns \textit{number of mates/neighbors} are provided in the \textit{``best\_neighbour''} cross-match tables with \gaia~DR3 from ESA for both \tmass~ and \allwise. 

\begin{figure*}[tp]
    \centering
    \includegraphics[width=\linewidth]{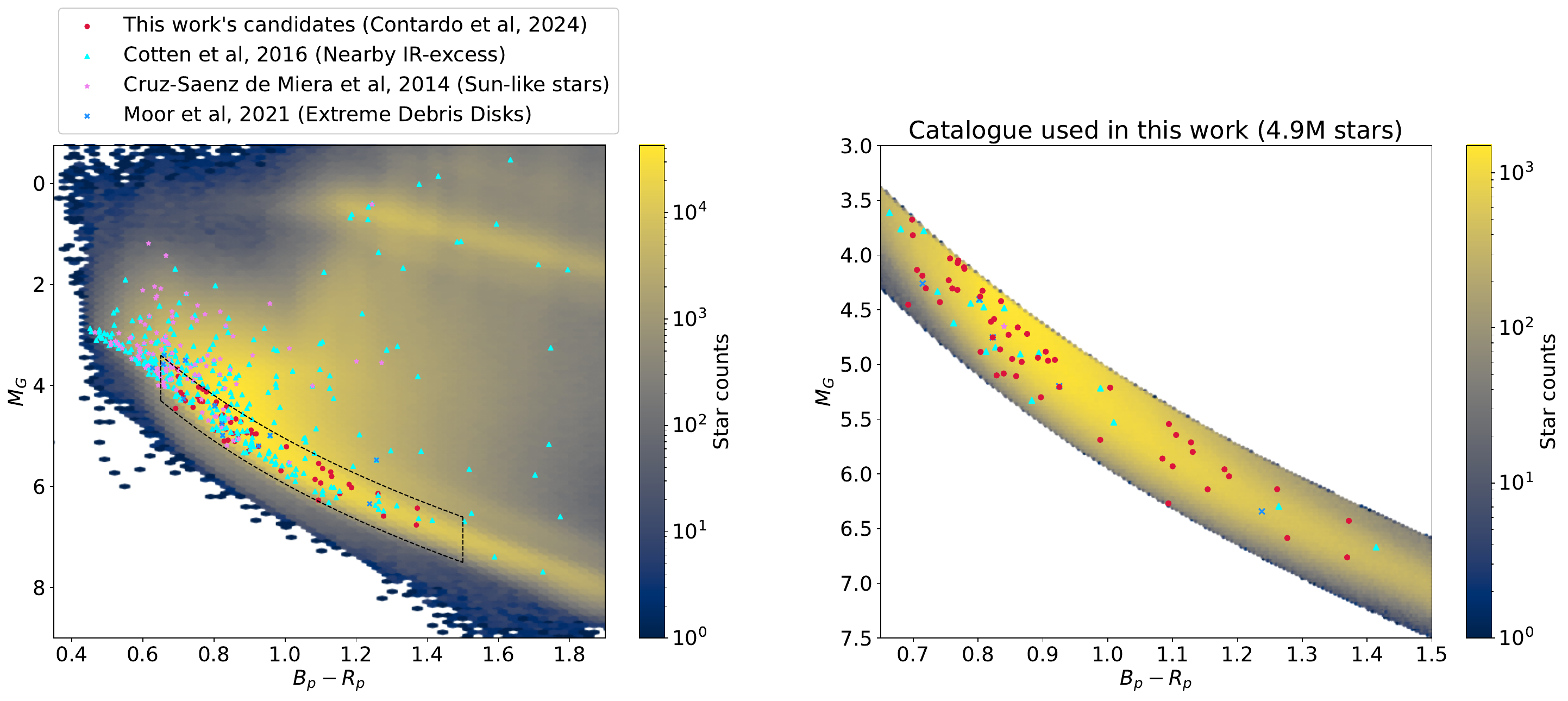}
    \caption{Left panel shows our initial sample of 18M FGK-type stars in \gaia{} DR3 with good cross-matches in \allwise{} and \tmass{}, in the color--absolute-magnitude space. The absolute magnitudes were computed as shown in Eq. \ref{eq:absmag}, and there were no corrections made for extinction or reddening by dust here. We plot IR-excess candidates from the literature that appear in this preliminary sample: from \cite{cotten2016census} (cyan, targeting nearby stars), \cite{cruz2014sunlike} (pink, targeting Sun-like stars), and \cite{moorEDD} (blue, targeting Extreme Debris Disks). Our final prime set of candidates is shown in red. The dashed black box illustrates our main-sequence cut.
    Right panel shows the final sample used for our search, after our main-sequence cut, dust cut and \unwise{} control cut as described in Section \ref{sec:data}. }    
    \label{fig:sample-HR}
\end{figure*}

Note that those cuts do not necessarily prevent potential contamination from close nearby sources if both sources were resolved in all catalogs (we suggest an additional crowding flag to address this in Section \ref{sec:method}). Finally, we impose that the angular distance between the \gaia~source and its best neighbors in the external catalogs is under $0.15$ arcseconds. 
The corresponding ADQL query for this cross-match selection is provided in the Appendix \ref{sec:appendix:adql}. It results in a preliminary dataset of 18,751,187 stars, or roughly about $1\%$ of Gaia DR3. The left panel of Figure \ref{fig:sample-HR} shows the distribution of this sample in the color-magnitude diagram.

We restrict this sample further by selecting stars along the main sequence, with the following color cut:
\begin{equation}
0.65 < B_p-R_p < 1.5
\end{equation}
and a cut in the color-absolute magnitude plane:
\begin{equation}
3.83 \times \log(B_p-R_p) + 5.05 < M_G < 3.83 \times \log(B_p-R_p) + 5.95 
\end{equation}

with $M_G$ computed as 

\begin{equation}
    M_G= G + 5 + 5 \times \log_{10}(\frac{\varpi}{1000\,\mathrm{mas}})
    \label{eq:absmag}
\end{equation}
where $\varpi$ is the measured parallax.

This reduces our dataset to 5,979,478 stars. 
We additionally apply a dust cut, using the \textsl{dustmaps} library \citep{2018dustmap} with the SFD \citep{1998SFD} dust reddening map. We discard sources with reddening values $E(B-V) > 1$. 

We then use the cross-match service provided by the \textsl{NOIRLab}'s \textsl{Astro Data Lab} to obtain the corresponding \unwise~values for $W1$ and $W2$ bands. We remove sources matched with more than one counterpart in the \unwise~cross-match table (where the original cone-search was of $1.5$ arc-seconds), and we apply a distance cut between matched sources under $0.15$ arc-seconds. We control for valid values in W1 and W2 magnitudes and restrict our sample to stars with \textit{unwise\_flags} values of $0$ for the bitmask for both $W1$ and $W2$ bands (i.e. sources are not deemed impacted by nearby bright sources or artifacts). 

This results in a final sample of 4,898,812 million stars.  The right panel of Figure~\ref{fig:sample-HR} shows the distribution of this sample in \gaia{} color-absolute magnitude diagram.

\section{Method}
\label{sec:method}

We propose to approach the problem of identifying infrared-excess stars in a data-driven fashion. Namely, our goal is to fit (or learn) a model on the data to predict the expected values for $W1$ and $W2$ bands, and to use the error in prediction as an indicator for anomalies, and potential infrared-excesses. Specifically, we are going to look for \textit{confidently wrong} predictions of the model.

We use the following features as input to our model: photometric observations (magnitude) from \gaia~DR3 and \tmass~ ($G,B_p,R_p$ mean magnitudes, and $J$, $H$, $K_s$), colors computed from \gaia~and \tmass~ ($B_p-R_p$, $B_p-G$, $G-R_p$, $J-H$, $J-K_s$, $H-K_s$, and $G-J$), and absolute magnitude $M_G$. We additionally use the \textsl{ruwe}, \textsl{parallax} and \textsl{ebpminrp-gspphot} features from \gaia. 

We fit independent Random Forest (RF) regressors to predict respectively the colors $K_s-W1$ and $K_s-W2$. We use the default setup from \textsl{sklearn} \citep{scikit-learn} using a mean-squared error criterion, $100$ trees, and no maximum depth. 
Comparisons with other (more flexible) models showed no improvement in prediction quality on a held-out sample, and therefore we chose this simple yet effective method.

\begin{figure}[tp]
    \centering
    \includegraphics[width=\linewidth]{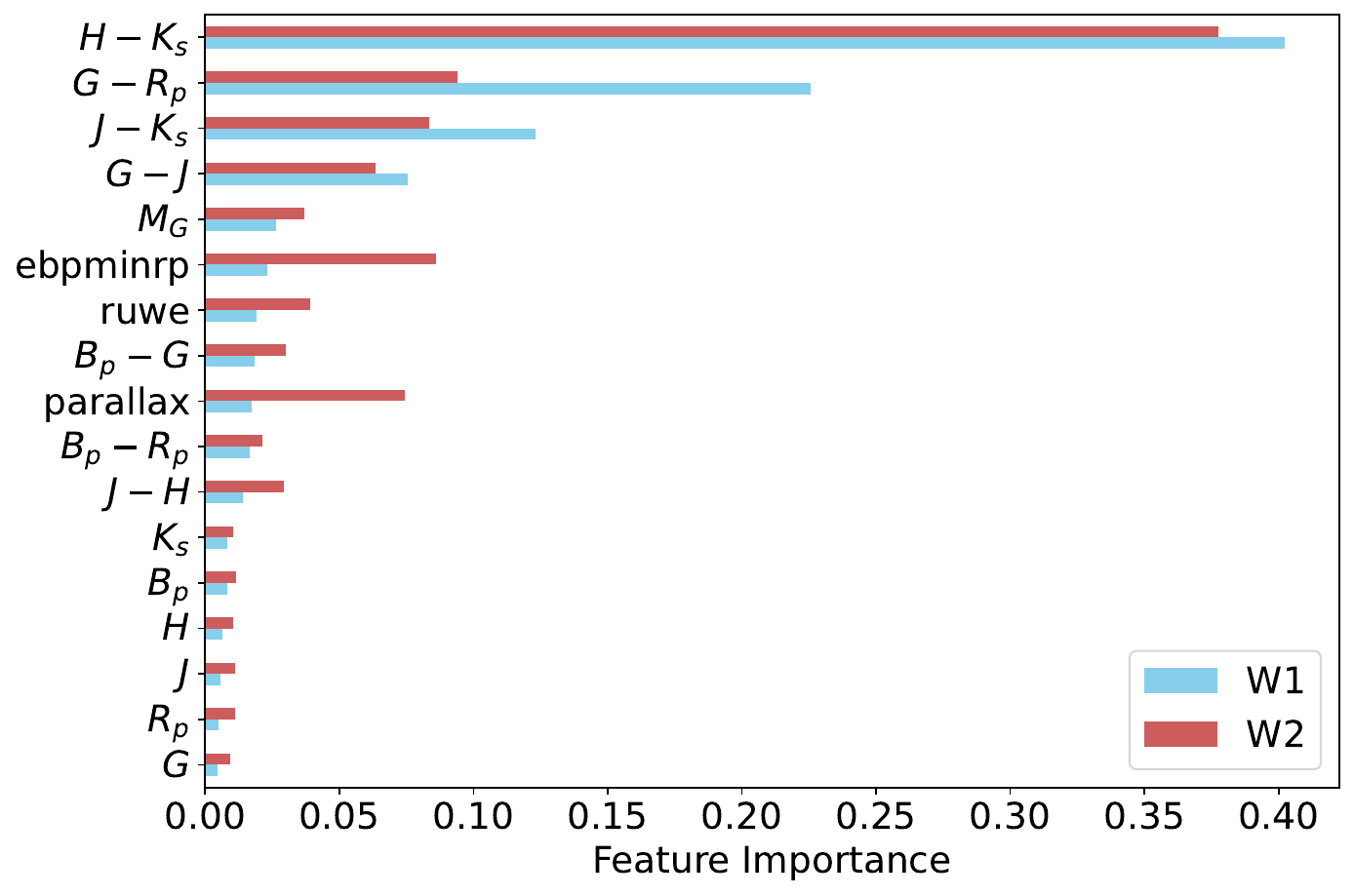}
    \caption{Feature Importance from two Random Forests fitted to predict $W1$ (blue) and $W2$ (red).}
    \label{fig:featureimportance}
\end{figure}

We perform an 8-fold split, where different RFs are trained separately on each fold ($12.5\%$ of the data or about 600,000 stars). We thus get 7 ``test'' predictions (for which the star did not appear in the training sample) for each star in our sample, for each color. Figure \ref{fig:featureimportance} shows the feature importances for a Random Forest predicting $K_s - W1$ (light blue) and an RF predicting $K_s - W2$. Interestingly, while colors features have high importance for both (as expected, since we predict a color), predictions of $W2$ seem to rely more on \textit{ebpminrp} (reddening value) and parallax than the RF for $W1$. The order of colors is also interesting: $G-R_p$ seem to have more importance than $J-H$ (likely because other colors from \tmass{} already have high importance).

From each of these 7 individual ``test'' prediction for color $K_s - W1$ (resp. $K_s - W2$), we can compute back the predicted $W1$ (resp. $W2$) magnitude of a star $i$, using the predicted color and the true observed $K_s$ magnitude of that star as: 
\begin{equation}
    \widetilde{W1}_{i,j} = Ks_{i} - RF1_j(x_i)
\end{equation}

Where $RF1_j(x_i)$ denotes the output (predicted color  $K_s - W1$) of the Random Forest $j$  for a star $x_i$. We can similarly compute a predicted $W2$ magnitude from the sets of Random Forest fitted to predict $K_s - W2$.

We can also compute an estimated magnitude $W1$ (resp. $W2$) for a star $i$ by combining the 7 test predictions, as the median prediction over the folds, which we denote: 

\begin{equation}
    \widehat{W}_i = \mathrm{Median}(\{\widetilde{W}_{i,j}\}_{j \in \mathrm{F}_i})
\end{equation}

Where $\mathrm{F}_i$ denotes the folds' indices where $i$ is not in the training set. Using $\widehat{W}_i$ as our predictor, we obtain a Mean Absolute Error of $0.016$ for $W1$ and $0.023$ for $W2$.

\begin{figure*}[tp]
\plottwo{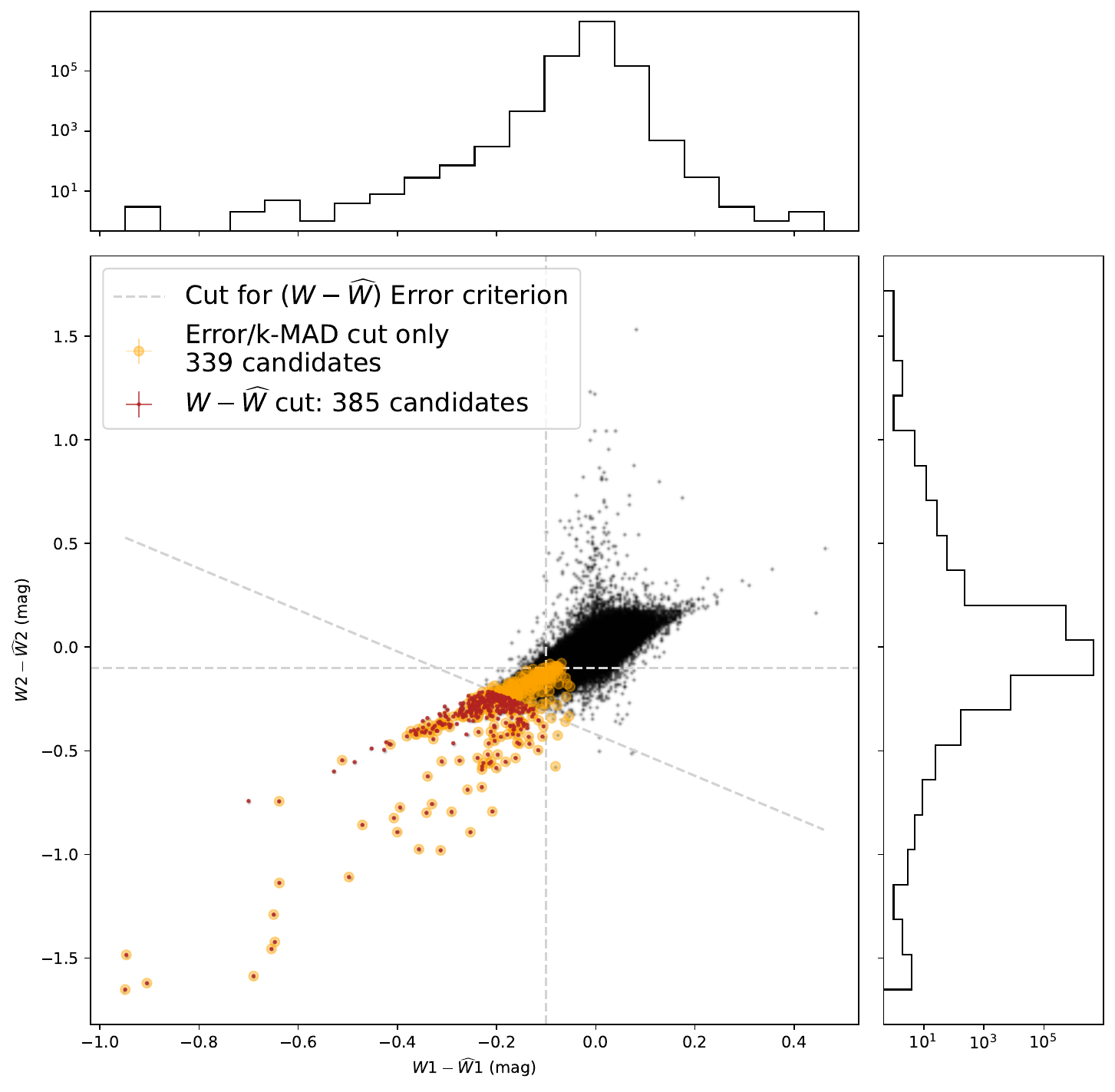}{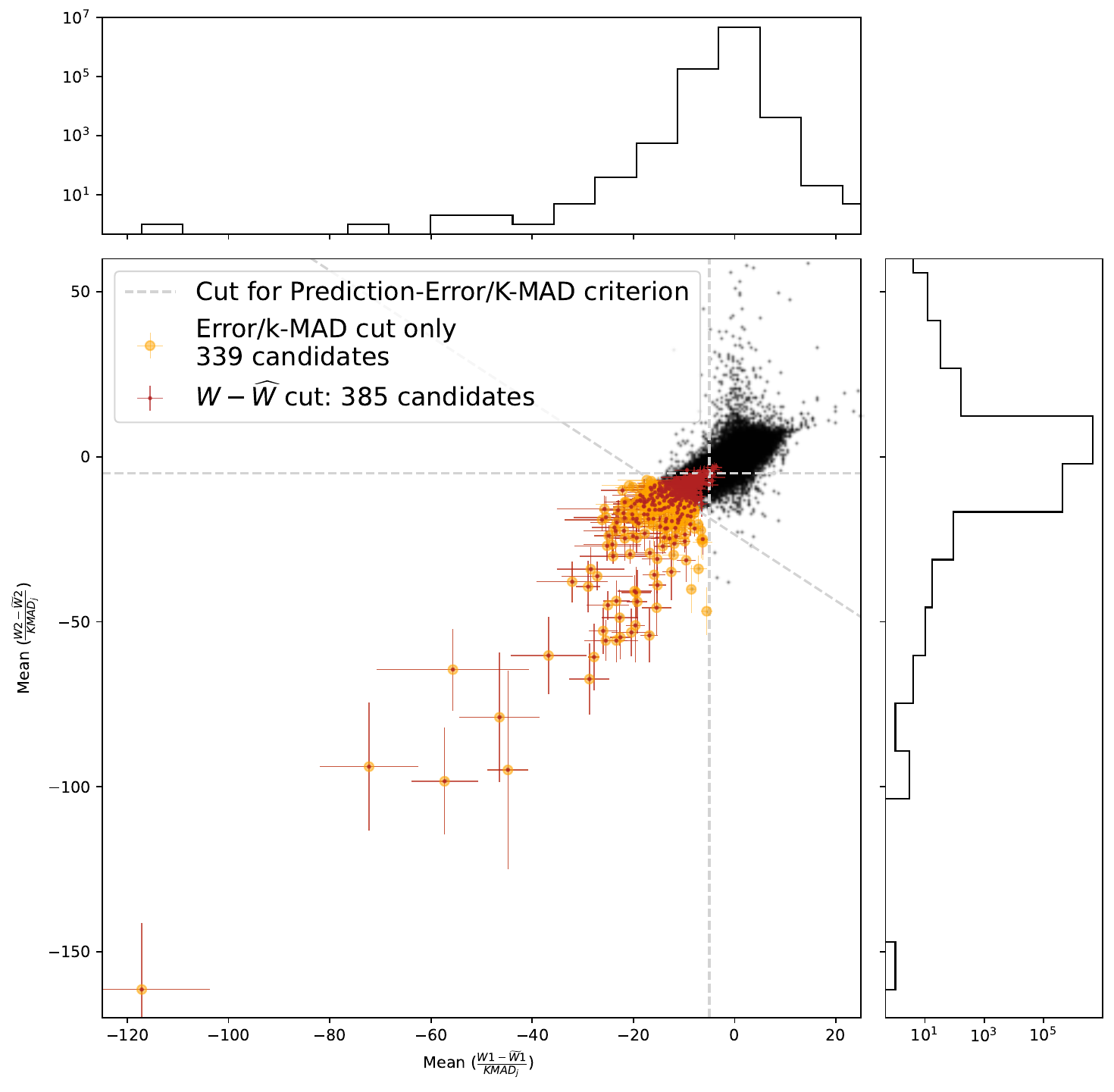}
    \caption{Left panel shows the difference between observed $W1$ (resp. $W2$ on $y$-axis) and the predicted $\widehat{W}1$ (resp. $\widehat{W}2$). We select stars with a high error in both bands as a preliminary sample of IR-excess candidates, depicted in red (cuts as dashed grey lines corresponding to Eq. \ref{eq:mean_error_cut}). The right panel shows the mean across 7-folds of the prediction error, divided by the Median Absolute Error of the object's 30 nearest neighbors in the respective fold. Error bars are standard deviations across the 7-folds. We select another set of candidates in this space, depicted in orange. Red data points match the ones in the left panel, and respectively the orange points match the ones in the right panel.}
    \label{fig:w1w2predscatter}
\end{figure*}

The left panel in Figure \ref{fig:w1w2predscatter} shows our sample in the $W - \widehat{W}$ error-space. Stars with very high prediction errors can be regarded as anomalous, either in excess (negative values) or deficit (positive values), as it means that the observed magnitude significantly deviates from the model prediction. Interestingly, one can see that significant infrared \textit{deficit} (positive values) in both bands (i.e. positive diagonal) are barely existent, as expected. We highlight as red points a set of preliminary candidates selected from a cut within that error space (dashed lines, corresponding to Equation \ref{eq:mean_error_cut} described below). We can also see in this Figure panel that there seem to be two ``types'' of excess candidates: one kind lying on the 1-to-1 line in terms of error in predicted magnitude, and another kind where the excess (error) is (roughly two times) stronger in W2, but less candidates lying in between those ``groups''.

While high prediction errors indicate potential anomalies, we also want to focus our search on highly confident incorrect predictions. Therefore, we propose to introduce additional criteria to flag candidates that (i) have a high prediction precision, i.e. the predictions across the folds have low variance (the error is not dependant on the training set),  (ii) are in \textit{well predicted} regions of the feature space (i.e. similar examples have a high accuracy), (iii) are in a well-populated region of the dataset (they are not outliers in the feature space).

\paragraph{Precision metric}: 
We define the fold-MAD (Median Absolute Difference) metric for a data point $i$ as the median of the absolute differences between each prediction from the RFs trained on different folds, and the global prediction $\widehat{W}_i$: 

\begin{equation}
\text{fold-MAD}_{i} = \mathrm{Median} (\{| \widehat{W}_{i} - \widetilde{W}_{i,j}|\}_{j \in \mathrm{F}_i})    
\end{equation}

This metric can be seen as a precision metric, relying on the scatter in the predictions for the object $i$.

\paragraph{Accuracy of similar examples}: For each object in our sample, we take their K-nearest neighbors ($K = 30$) in the following feature space, in each test-fold: $H-K_s, G-R, J-K_s, G-J, B-G, M_G$, parallax, \textit{ebpminrp} (reddening), and ruwe. (Those features have been identified as the ``most important'' features for predictions by the Random Forests). The features are rescaled in a MinMax fashion before computing the neighbors. We refer to this KNN as NN-color. We compute the K-MAD of a star $i$ in the test-set of fold $j$ as the median of the absolute differences between the true and predicted magnitudes $\widetilde{W}_{k,j}$ over the $K$ neighbors in that test-set:

\begin{equation}
     \text{K-MAD}_{i,j} = \mathrm{Median} (\{|W_{k} - \widetilde{W}_{k,j}|\}_{k \in \text{NN-colour}(i,j)})
\end{equation}
Where $\text{NN}(i,j)$ denotes the $K$ nearest neighbors indices for the object $i$ in the test-set of fold $j$, and $W$ denotes either $W1$ or $W2$ bands. In other words, this measures if similar examples to a given object are well predicted or not. 
We can use this to favor candidates that have high errors but a low K-MAD (i.e. candidates for which similar objects were correctly predicted), increasing our confidence in the anomalous nature of the object, or to reweight the significance of an error in prediction through the K-MAD (i.e. a small prediction error is ``more significant'' if the neighbors are very accurately predicted). We note however that if we had a ``cluster'' of, say,  objects with an IR excess that were all very similar in the feature space (enough so that they'd form the majority of each others' neighborhood), this criterion could mechanically discard these objects. We consider however that, given the size of our sample, the feature space we look at and the rarity of the objects we are looking for, this is relatively unlikely to happen.

The right panel of Figure \ref{fig:w1w2predscatter} shows the distribution of our sample in the prediction error divided by the K-MAD (computed per fold), averaged across the test-folds, for $W1$ and $W2$. We also plot the standard deviation across the folds as error bars for a selection of infrared excess candidates (negative values) under this criterion, shown in orange. The red candidates (high prediction errors) from the left panel of the same Figure match the red points in the right panel, and vice-versa for orange points, showcasing how some high-error sources might be significantly down-weighted by the use of the K-MAD. 

\paragraph{Well-populated region of the feature space}: Additionally, we can compute the distance between the source and its neighbors. We can use this metric as a criterion for flagging or removing candidates that might live in a low-density region of the input feature space (hence with possibly less reliable prediction, and for which the K-MAD might not be indicative).

We propose to further investigate a set of candidate stars that are selected through a combination of the prediction error criterion (red points in Figure \ref{fig:w1w2predscatter}), the error over K-MAD criterion (orange point in Figure \ref{fig:w1w2predscatter}, down-weighting the points that are in a region of the feature-space that is badly predicted, favoring high-accuracy regions), and the fold-MAD, to ensure our candidates have small prediction uncertainties (low scatter). 
More specifically, we select the stars with an error in each band under the $0.008$-th percentile value (dashed light-grey diagonal line in the left panel of Figure \ref{fig:w1w2predscatter}), and with both errors under $- 0.1$ mag (horizontal and vertical dashed light-grey line): 

\begin{equation}
\begin{split}
    & (W1_i - \widehat{W1}_i) + (W2_i - \widehat{W2}_i) < -0.42 \\
& (W1_i - \widehat{W1}_i) < -0.1 \\
& (W2_i - \widehat{W2}_i) < -0.1
\end{split}
\label{eq:mean_error_cut}
\end{equation}

For the error over k-MAD cut, we similarly select stars with a sum of the mean of error over k-MAD under the $0.007$-th percentile value, and with both means of error over K-MAD under $-5$:
\begin{equation}
\begin{split}
    &\frac{1}{7}\sum_{j \in \mathrm{F}_i} \frac{(W1_i - \widetilde{W1}_{i,j})}{\text{K-MAD1}_{i,j}} + \frac{1}{7}\sum_{j \in \mathrm{F}_i}\frac{(W2_i - \widetilde{W2}_{i,j})}{\text{K-MAD2}_{i,j}} < -23.59 \\
& \frac{1}{7}\sum_{j \in \mathrm{F}_i} \frac{(W1_i - \widetilde{W1}_{i,j})}{\text{K-MAD1}_{i,j}} < -5 \\
& \frac{1}{7}\sum_{j \in \mathrm{F}_i}\frac{(W2_i - \widetilde{W2}_{i,j})}{\text{K-MAD2}_{i,j}} < -5
\end{split}
\label{eq:mean_errorMAD_cut}
\end{equation}

Which corresponds to the dashed grey lines in the right panel of Figure \ref{fig:w1w2predscatter}.
\\We additionally select candidates that have a fold-MAD under $0.01$ mag:

\begin{equation}
\begin{split}
    &    \mathrm{Median} (\{| \widehat{W1}_{i} - \widetilde{W1}_{i,j}|\}_{j \in \mathrm{F}_i})     < 0.01 \\
    & \mathrm{Median} (\{| \widehat{W2}_{i} - \widetilde{W2}_{i,j}|\}_{j \in \mathrm{F}_i})     < 0.01
\end{split}
    \label{eq:foldmad_cut}
\end{equation}

This results in a preliminary set of excess candidates of 127 stars.

We note that this threshold cut is rather arbitrary: we naturally focus on a few numbers of extreme outliers and enforce a relationship in the errors in both $W1$ and $W2$. Other strategies could be used to decide where to put this threshold. For instance, one could use the rate or contamination in the symmetric, infrared-deficit side: such events should not occur, and thus this gives us some information about the ``scatter'' in our data and predictions and possible false-detection rate in our candidates (see Section \ref{sec:occurencerate} and Appendix \ref{sec:appendix:deficit} for more details).

\subsection{Additional cuts to reduce false-detection by contamination and other artifacts}

On this preliminary set of 127 candidates, we evaluate the following additional features that could lead to false anomalies:

\begin{itemize}
    \item \textbf{Crowding:}  For each object, we perform a cone-search in \gaia{}, \tmass{}, and \unwise{} catalogues of $5$ arcseconds. If another object is found within that cone for any of the catalogs, we flag it as potentially being contaminated, or in a crowded area. 
    \item \textbf{Figure of Merit (FoM) from \gaia}: Following the recommendations in \cite{2020wordtowise} to alleviate source confusion when looking for IR-excess in \wise, we check the figure of merit (\textit{score}) provided by ESA \gaia{} cross-match with \allwise. This score assesses the quality or ``likeliness' of the match (the higher the better), however, it is not normalized. \cite{2020wordtowise} suggest as a rule of thumb to discard sources with a high SNR but a low FoM (they suggest that excesses with SNR $> 10$ but FoM $< 4$ are likely the result of source confusion). We do not have objects with a FoM under $4$ in our preliminary set of candidates. 
    \item \textbf{Proper-Motions Comparison}: \cite{2020wordtowise} also suggest to check proper motions agreements between catalogs to remove potential source confusion. They indicate that the proper motion accuracy in \textsl{CatWise} depends on the magnitude of the source ($10$ mas yr$^-1$ for bright sources, $30$ mas yr$^-1$ for $W1 \approx 15.5$ mag). We compare the proper motions from \gaia{} and from \textsl{CatWise} and apply a cut when the proper motions in either $RA$ or $DEC$ disagree more than the approximated accuracy for the source brightness. 
    \item \textbf{Disagreement between \unwise{} and \allwise}: To ensure confidence in the observed magnitude of our targets, we add a flag checking for agreement between \unwise{} and \allwise. \unwise{} documentation recommends subtracting 4 mmag from unWISE W1 and 32 mmag from unWISE W2 to improve agreement between both catalogues. After doing so, we add an ``agreement-flag'' such that $|W^{\allwise} - W^{\unwise}| < 0.15$ for both bands.
    \item \textbf{Distance to galactic plane}: We remove candidates that are close to the galactic plane s.t. $abs(b)<10$.
    \item \textbf{Binary and duplicated sources}: Candidates that are indicated as ``non single stars'' by \gaia{} DR3 are flagged. We also check each candidate with Simbad (using \gaia{} DR3 ID) and flag each star that is indicated to be eclipsing binaries or eclipsing binaries candidates. \gaia{} DR3 also provides a \textit{duplicated source} flag (which may indicate observational, cross-matching or processing problems, stellar multiplicity, and possible unreliable astrometry or photometry). We therefore discard candidates with this flag.
\end{itemize}

We combine these different flags to obtain a curated final set of $53$ MIR-excess candidates. We provide in Table \ref{tab:cuts} the number of candidates remaining after applying each cut. The \gaia{} DR3 and \tmass{} ID of those 53 candidates are available in Table \ref{tab:candisummary} in the Appendix \ref{sec:appendix:canditable}, and as a csv file at \url{https://github.com/contardog/NotATechnosignatureSearch} along with the code for this pipeline and the following analysis.

\begin{deluxetable*}{cc}[tp]
\tablecaption{Table of criterion for our candidates' selection, and the number of candidates remaining after each cut.
\label{tab:cuts} }
\tablehead{\colhead{Criterion cut} & \colhead{Number of remaining candidates} }
\startdata
Prediction Error cut (Eq. \ref{eq:mean_error_cut}) & 385 \\ 
Mean (Prediction Error / k-MAD) cut (Eq. \ref{eq:mean_errorMAD_cut}) & 339 
\\ \hline
Error cut AND error/k-MAD cut & 170 \\
fold-MAD cut (Eq. \ref{eq:foldmad_cut}) & 127 \\
Crowding cut at 5 arcsecond & 87 \\
FoM $> 4$ cut & 87 \\
Proper-Motion disagreement cut & 78 \\
Disagreement \allwise{}/\unwise{} cut & 76 \\
Mean Distance k-NN $< .1$ & 66 \\
$abs(b) > 10$ & 59 \\
Removing binaries and binaries candidates (\gaia{}, Simbad) & 55 \\
Removing duplicated sources (\gaia{} DR3 flag) & 53 \\
\enddata
\end{deluxetable*}

\section{Analysis of Excess Candidates}
\label{sec:analysis}

\subsection{Characterization of the hosts}
\begin{figure}
    \centering
    \includegraphics[width=\linewidth]{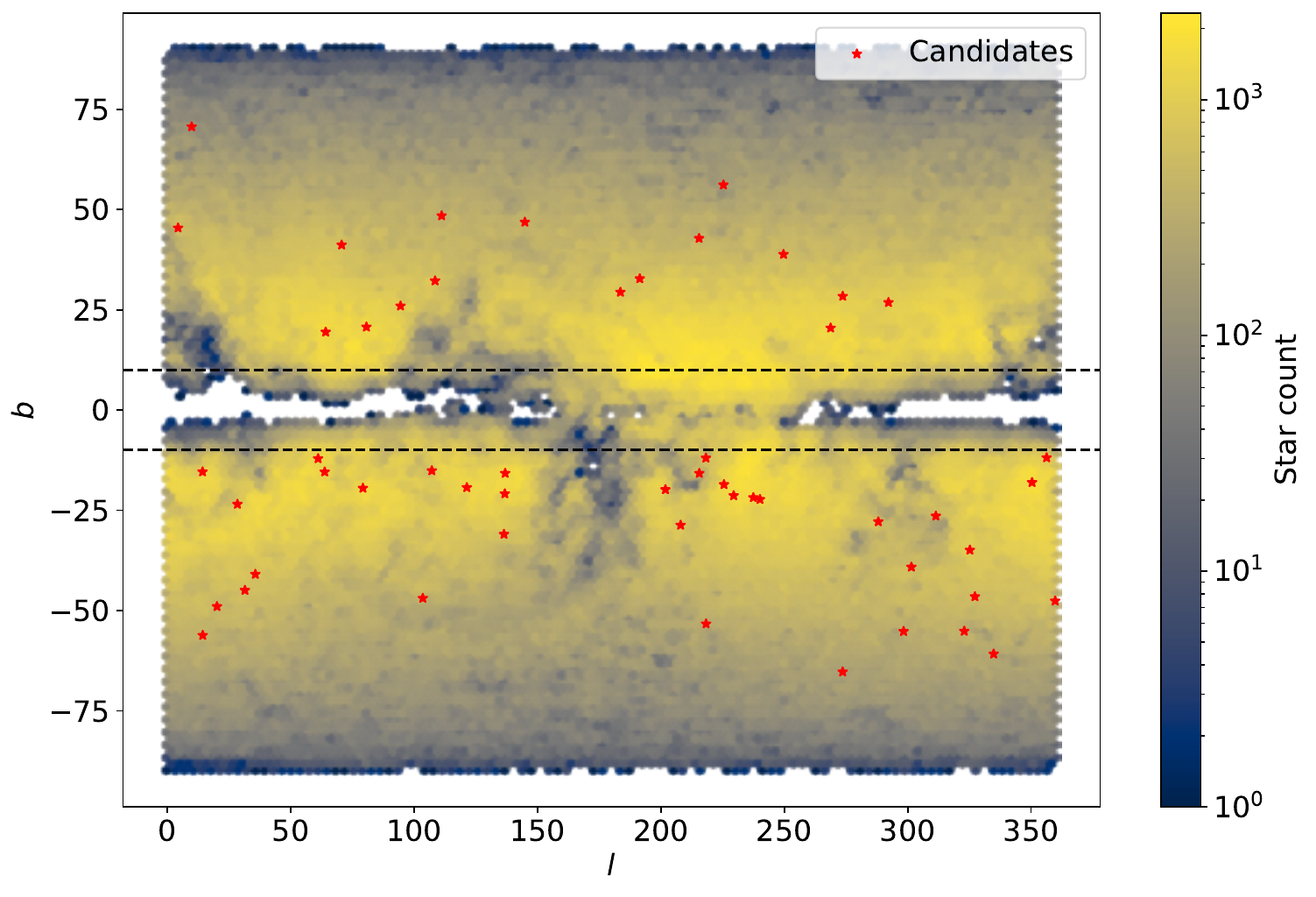}
    \caption{Distribution of our full sample and our candidates in Galactic coordinates. Dashed black lines indicate $abs(b)=10$. }
    \label{fig:galcandidate}
\end{figure}

\begin{figure*}
    \centering
    \includegraphics[width=.8\textwidth]{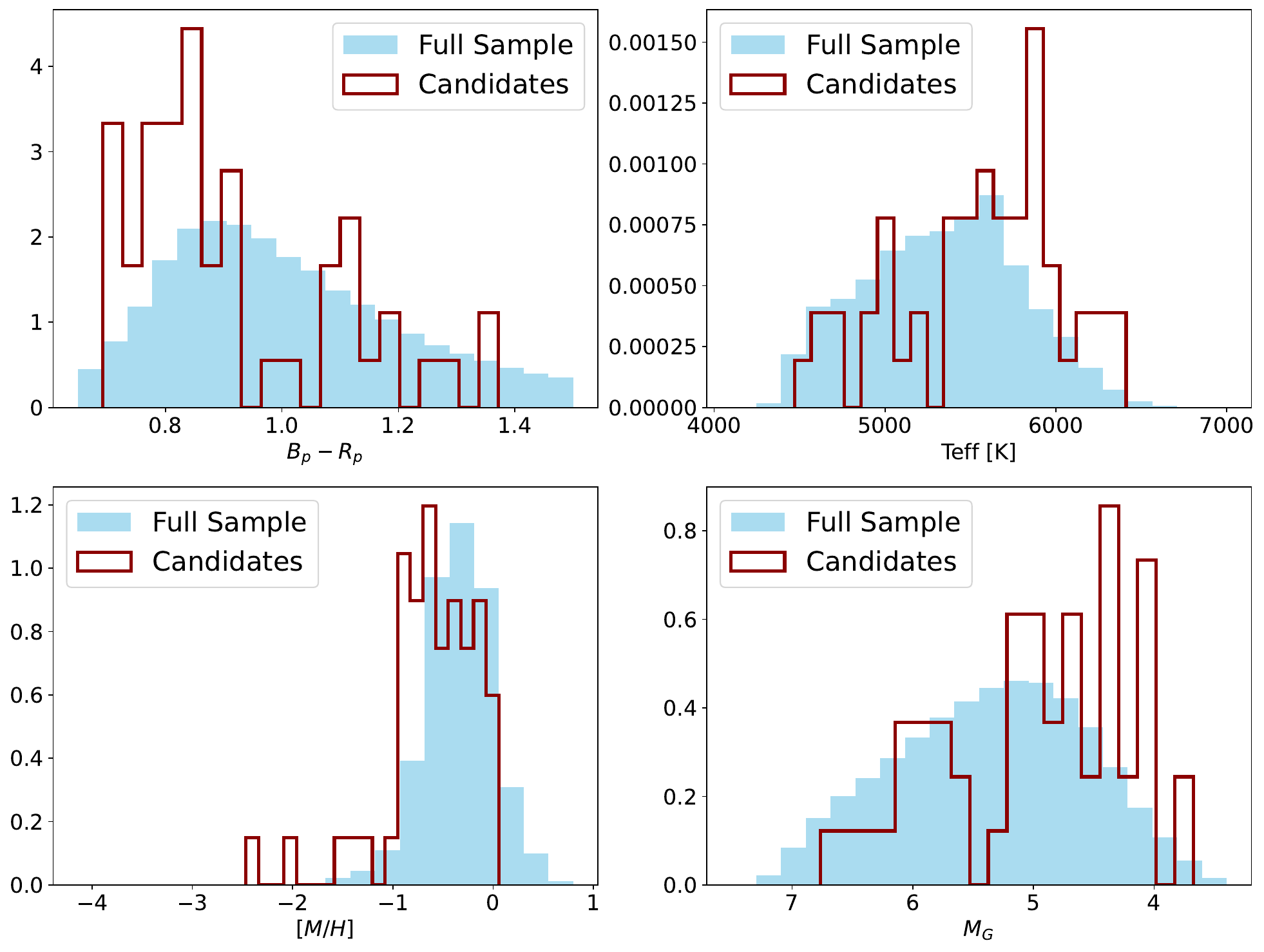}
    \caption{Histograms of distribution of our candidates' sample (dark red outline) and our full 4.9M stars sample (blue) in $B_p-R_p$, effective temperature, $[M/H]$, and $M_G$.}
    \label{fig:candi_hist}
\end{figure*}

Figure \ref{fig:galcandidate} shows where our candidates lie in Galactic coordinates $l,b$, where the dashed lines indicate the cut at $abs(b) < 10$.  We provide in Figure \ref{fig:candi_hist} histograms of our candidates' sample and of our complete sample in $B_p-R_p$, $[M/H]$ (obtained from \gaia~DR3 \textit{mh\_gspphot}), Teff, and $M_G$. We observe a gap in our candidates' distribution around 1 in $B_p-R_p$, 5.5 in $M_G$ and $5400$K in Teff. There also seems to be an over-density in our candidates' distribution compared to the full sample of stars with color $B_p-R_p < 0.9$ and Teff $\sim 5900$K. Because of the data-driven nature of our approach, and the subsequent cuts used to create our candidates' sample, as well as the small size of our final sample, it is hard to give clear interpretations and conclusions about these trends. We note however that the dip observed here at $B_p-R_p \sim 1$ is not strongly visible when looking at the distribution of the candidates selected using the Prediction Error only (Eq. \ref{eq:mean_error_cut}) or the $k$-MAD cut only (Eq. \ref{eq:mean_errorMAD_cut}), and only start appearing when combining Eqs \ref{eq:mean_error_cut}, \ref{eq:mean_errorMAD_cut} and \ref{eq:foldmad_cut} altogether, and is further depleted by subsequent cuts (at a somewhat similar rate than other color-bins). We also observe an offset trend in $M/H$ towards lower values than the full sample, with a few outliers standing out. 

For the four observables in Figure \ref{fig:candi_hist}, we conducted a Kolmogorov-Smirnoff test which indicates that the differences in distribution with the full sample are significant (with p-values $< 1e^{-3}$).

\subsection{H-alpha emission}
Our sample focuses on the main sequence, but stars might have only recently joined and could be in a ``transitional'' phase between protoplanetary disks and debris disks \citep{wyatt2015evoldisks}. To properly understand the nature(s) of the excesses observed in our candidates, and to identify the most interesting candidates for follow-up observations (e.g. confidently old stars with an unusually high IR excess), it is crucial to constrain the hosts' ages. We investigate different markers that could indicate that the host is young or recently joined the main sequence. However, the analysis presented below is only a first step and we defer thorough age-estimations of the candidates to future work.

\begin{figure*}[tp]
    \centering
    \includegraphics[width=\linewidth]{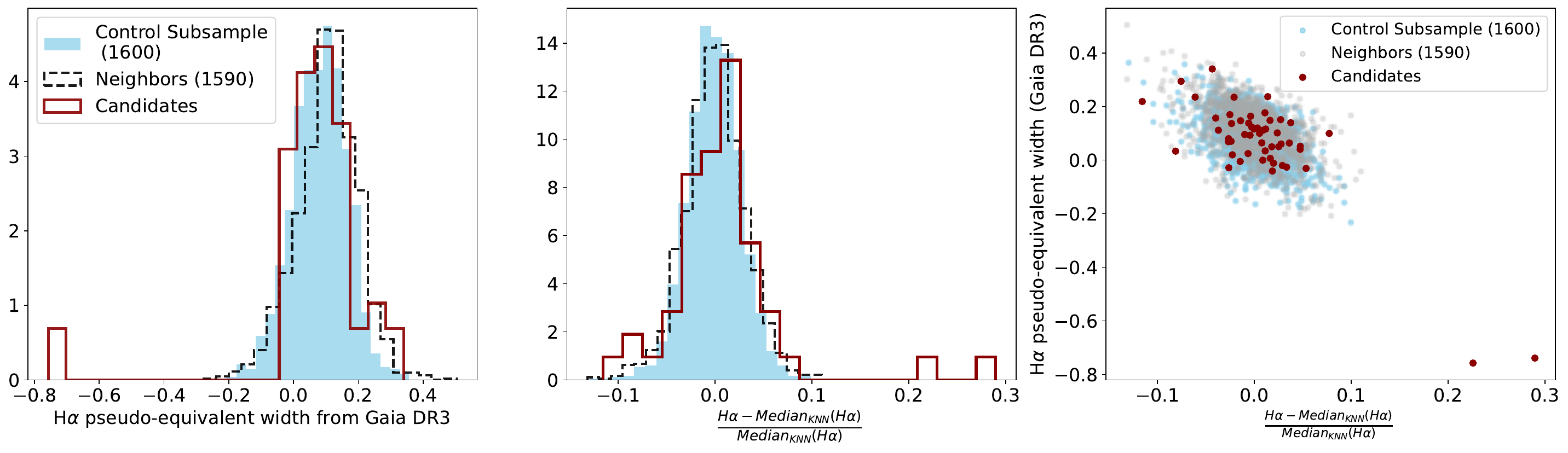}
\caption{Distribution of measurements for H$\alpha$ emission for a random control subsample (blue), a set of stars similar to our candidates (black dashed line) and our candidates set (dark red line). Left panel shows the distribution of the pseudo-equivalent width of the H$\alpha$ line provided by \gaia{} DR3 (negative values indicate emission). Middle panel shows our measurement using Equation \ref{eq:halpha} (positive values indicate emission). Right panel shows those measurements against each other.}
\label{fig:halpha_triple}
\end{figure*}

To investigate the potential ``youngness'' of our candidates, we start by evaluating the H$\alpha$ emission. We first use the measurement of the pseudo-equivalent width of the H$\alpha$ line provided by  \gaia{} DR3 from the RP spectra, \textit{ew\_espels\_halpha}. The value is expected to be negative when emission is present. The left panel of Figure \ref{fig:halpha_triple} shows the distribution of this measurement for our candidates' sample (dark red line), as well as a randomly selected control sample of 1,600 stars from our 4.9M sample (blue) and a set of stars similar to our candidates called ``Neighbors'' (dashed black line). This set of neighbors is computed in a different features-space than previously. We use the following features to favor stars of similar magnitude and spectral shape (while our previous KNN relied on color): $M_G$, \textit{ebpminrp\_gspphot, parallax, ruwe}, $G$, $J$, $H$, $K_s$, $B_p$, and $R_p$. We refer to this KNN as NN-mag. 

We can see two clear outliers in our set of candidates with high H$\alpha$ emission, indicating young stellar objects. The rest of our candidates seem to follow the control distribution, perhaps with a slight over-density on the absorption side. The pseudo-equivalent width EWH$\alpha$ of all candidates is given in the summary Table \ref{tab:candisummary} in Appendix \ref{sec:appendix:canditable}.

As a sanity check, we perform a different measurement of potential emissions or absorptions of H$\alpha$ as follows: using the $XP$ spectra and \textsl{GaiaXPy} library, we measure the flux value of the spectra at the H$\alpha$ wavelength (656.46nm). For an object $i$, we obtain the flux values of its 30 nearest neighbors (using NN-mag), and compute the difference between the object's flux and the median flux of its neighbors, divided by the median flux:

\begin{equation}
    E_{H\alpha, i} = \frac{F_{H\alpha, i} - \text{Median}_{k \in \text{NN-mag}(i)}(F_{H\alpha,j})}{\text{Median}_{k \in \text{NN-mag}(i)}(F_{H\alpha,j})}
\label{eq:halpha}
\end{equation}

This metric provides an indicator of how much the object flux diverges at the H$\alpha$ line from similar objects, where a positive value would indicate emission (respectively absorption for negative values). The middle panel of Figure \ref{fig:halpha_triple} shows the distribution of $E_{H\alpha}$ for our set of candidates, the control sample and the neighbors sample, and the right panel shows $E_{H\alpha}$ against the pseudo-width provided by \gaia{} data product. We see agreement especially in the detection of the two high emissions objects, \gaia{} DR3 5260430146906778624 (top right panel in Figure \ref{fig:SED1}) and \gaia{} DR3 3004530769758177280 (third row, left panel of Figure \ref{fig:SED2}). Note that the second object corresponds to our candidate with the highest fractional luminosity ($\sim 0.1$).

Visualizing the $XP$ spectra also highlights some odd objects with respect to their NN-mag neighbors (in grey), see for instance the top middle candidate in Figure \ref{fig:SED0} (\gaia{} DR3 6705842630930258816). 

\begin{figure}
    \centering
    \includegraphics[width=\linewidth]{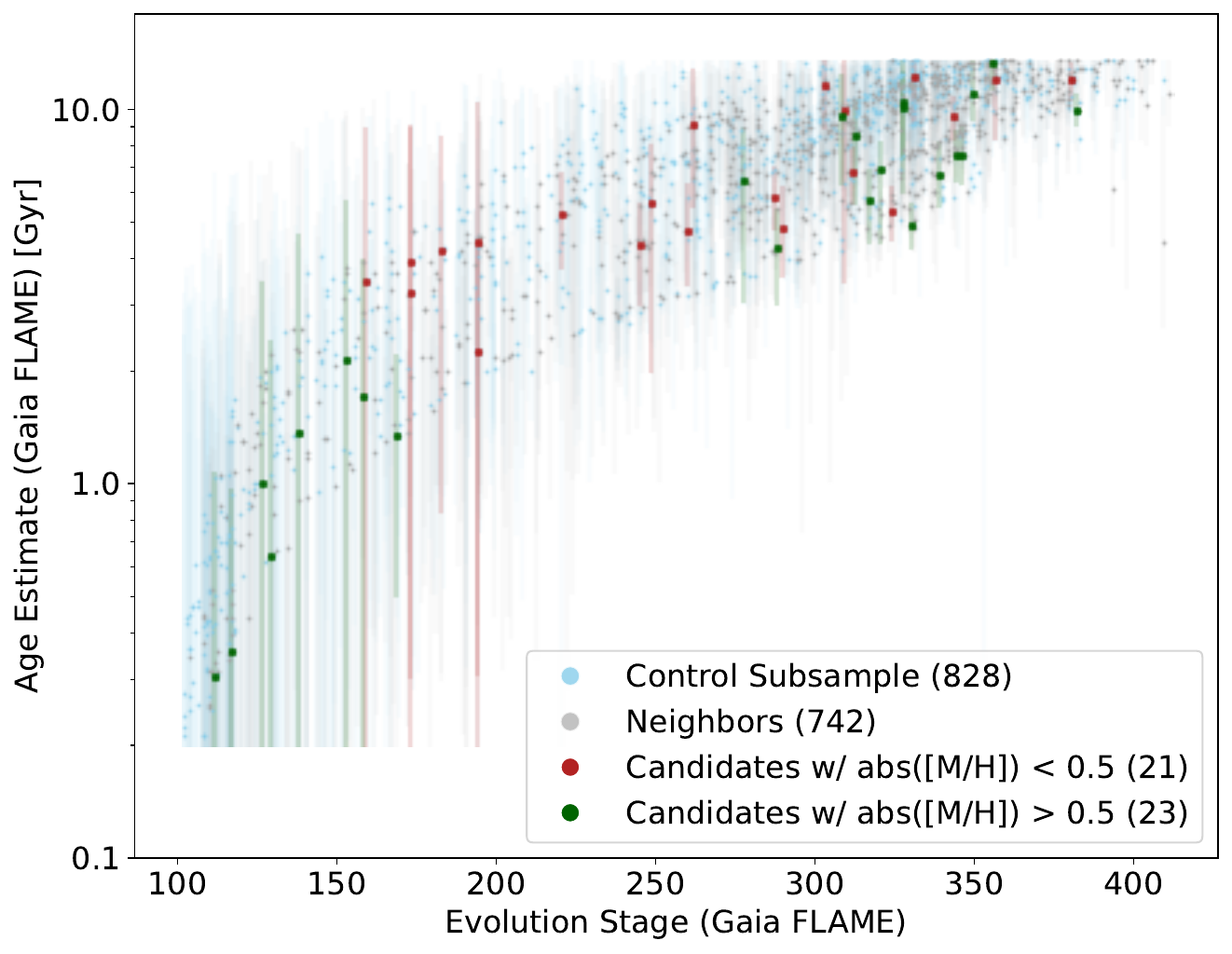}
    \caption{Age estimates and evolution stage provided by the Final Luminosity Age Mass Estimator (FLAME) in \gaia{} DR3, for our candidates (dark red), a set of similar stars (grey) and a control sample (blue), restricted to the stars that have $-0.5 <  [M/H]  < +0.5$. The error bars depict the lower (16th percentile) and upper (84th percentile) ages to cover the $68\%$ confidence interval. The $x$-axis Evolution Stage goes from 100 (Zero Age Main Sequence) to 490 (Red Giant Branch). 360 would be the main sequence turn-off. }
    \label{fig:ageflame}
\end{figure}

For the sake of completeness and using the data product available with the catalogs we used, we also provide in Figure \ref{fig:ageflame} a visualization of the age estimates provided by the Final Luminosity Age Mass Estimator (FLAME) within \gaia{} \citep{creevey2023gaiaApsis} for our candidates and control samples. The ``error bars'' correspond to the lower and upper confidence levels of the FLAME's age estimates (i.e. the 16th and 84th percentile value of the 1D projected distribution from sampling in mass and age). The $x$-axis is the ``evolution stage'' from FLAME, where 100 corresponds to Zero Age Main Sequence (ZAMS), 360 to the main sequence turn-off, and 490 would be the base of the Red Giant Branch \citep{hidalgo2018basti}. However, \cite{creevey2023gaiaApsis} highlights that those age estimates assume some prior on the metallicity, and the authors advise caution in using those ages for stars outside of the $-0.5 <  [M/H]  < +0.5$ regime. Therefore, we plot only in Figure \ref{fig:ageflame} the stars that have a metallicity estimate (provided by \gaia{}) in this range for the control and neighbors subsample. For our candidates, we plot in red the stars within the correct metallicity range, and in green the ones outside this range. All the youngest ($<$2Gyrs) candidates fall outside of the metallicity range. However, some of the remaining candidates have large error bars. One of the candidates with a high H$\alpha$ emission has an age estimate of 5.79 Gyrs (the other candidate does not have an estimate). We therefore recommend cautiousness in making any conclusion from those estimates. 

~\\~\\
\subsection{Check for Radio sources}
Following \cite{ren2024hephaistosradio}, we search for cross-matching radio sources with our candidates: radio sources in the close vicinity of a candidate might indicate possible contamination with background galaxies (e.g. dust obscured). We search for radio sources within 15" of our candidates in catalogs from the Very Large Array Sky Survey (VLASS, \cite{gordon2021radioVLASS}), Rapid ASKAP Continuum Survey (RACS, \cite{hale2021RACSradio}), the FIRST survey \citep{helfand2015FIRSTradio}, the NRAO VLA Sky Survey (NVSS, \cite{condon1998VLAradio}), the TIFR GMRT Sky Survey (TGSS, \cite{intema2017TGSSRadio}), and we add to this list the LOFAR Two-metre Sky Survey (LoTSS) DR2 \citep{shimwell2022LOFARradio}. 

We found radio sources near 6 of our candidates. Four of those have a radio source within 1"; one within 2", and one at 10."65. A flag indicating candidates with a nearby radio source is included in the summary Table \ref{tab:candisummary} provided in the Appendix \ref{sec:appendix:canditable}.


\subsection{Variability of hosts}

Another matter of interest in sources of IR excess is the variability of the object. Following \cite{moorEDD}, we use the \textsl{NeoWISE} time series to investigate first the variability of our candidates in the mid-infrared. We use the correlation-based Stetson index \citep{stetson1996} $S_J$, which measures the correlation of the variability in two (or more) bands and can be computed for a given object as follows:
\begin{equation}
    S_J = \frac{\sum_{j=1}^n w_k \text{sgn}(P_k)\sqrt{|P_k|}}{\sum_{j=1}^n w_k}
    \label{eq:stetson}
\end{equation}
Where $n$ is the number of paired observations across the different bands, $w_k$ is a weight per epoch (uniform in our case) and :
\[P_k = (\sqrt{\frac{n}{n-1}} \frac{W_{1,k} - \bar{W}_1}{\sigma_{W_{1,k}}} )(\sqrt{\frac{n}{n-1}} \frac{W_{2,k} - \bar{W}_2}{\sigma_{W_{2,k}}} )\]
Where $\sigma_{W_{i,k}}$ is the uncertainty for the $k$-th observation in the $i$-th band, and $\bar{W}_i$ is the mean of the observed points in band $i$.

\begin{figure}
    \centering
    \includegraphics[width=\linewidth]{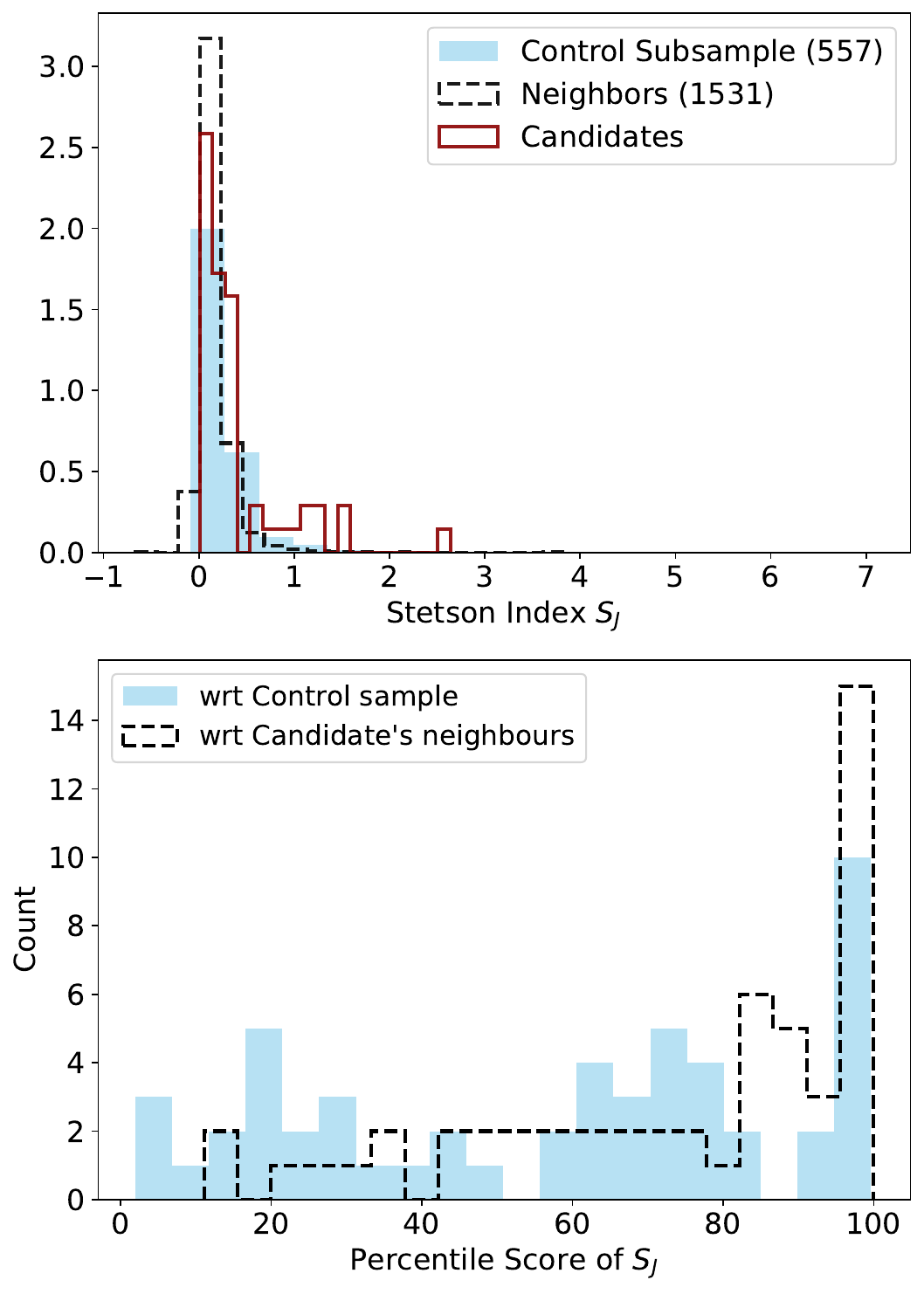}
    \caption{Left panel shows the distribution of the Stetson index (Equation \ref{eq:stetson}) of our control sample (blue), of a set of stars similar to our candidates (black dashed line) and of our candidates (dark red line). Right panel shows the distribution of the percentile score of our candidates' Stetson index with respect to the control sample (blue) and with respect to their set of similar stars (black dashed line).}
    \label{fig:stetson}
\end{figure}

\begin{figure*}[t]
\plottwo{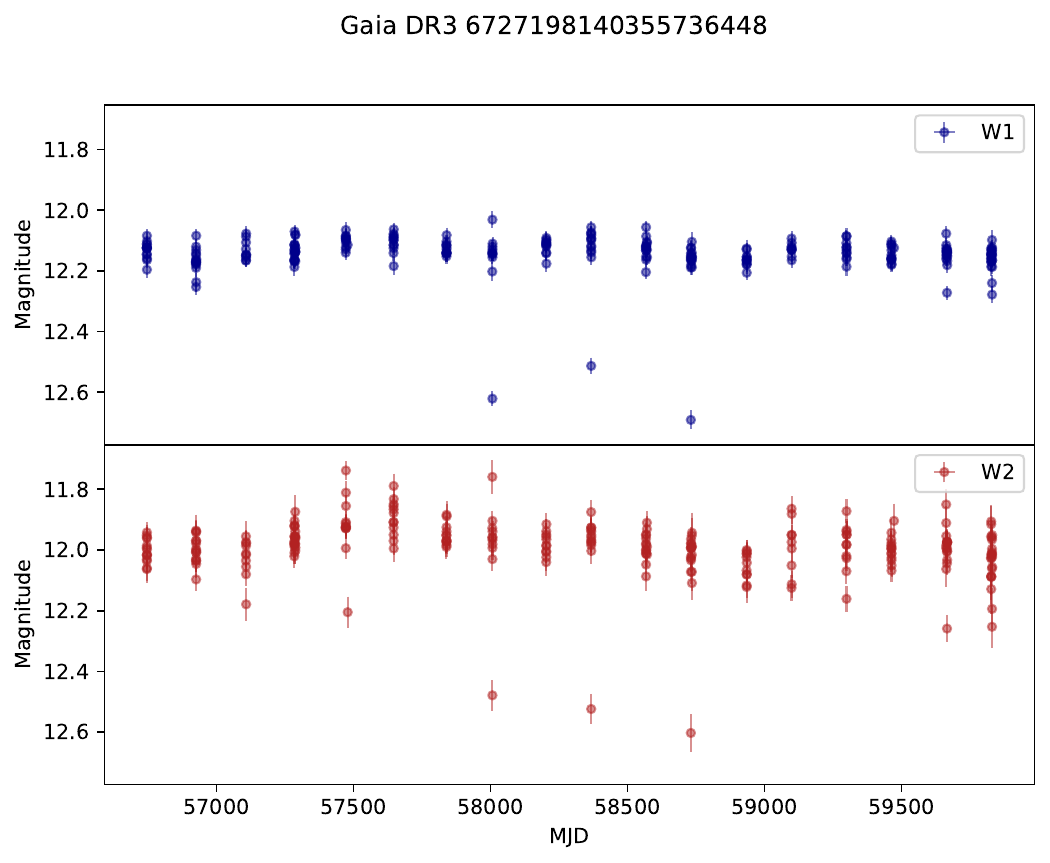}{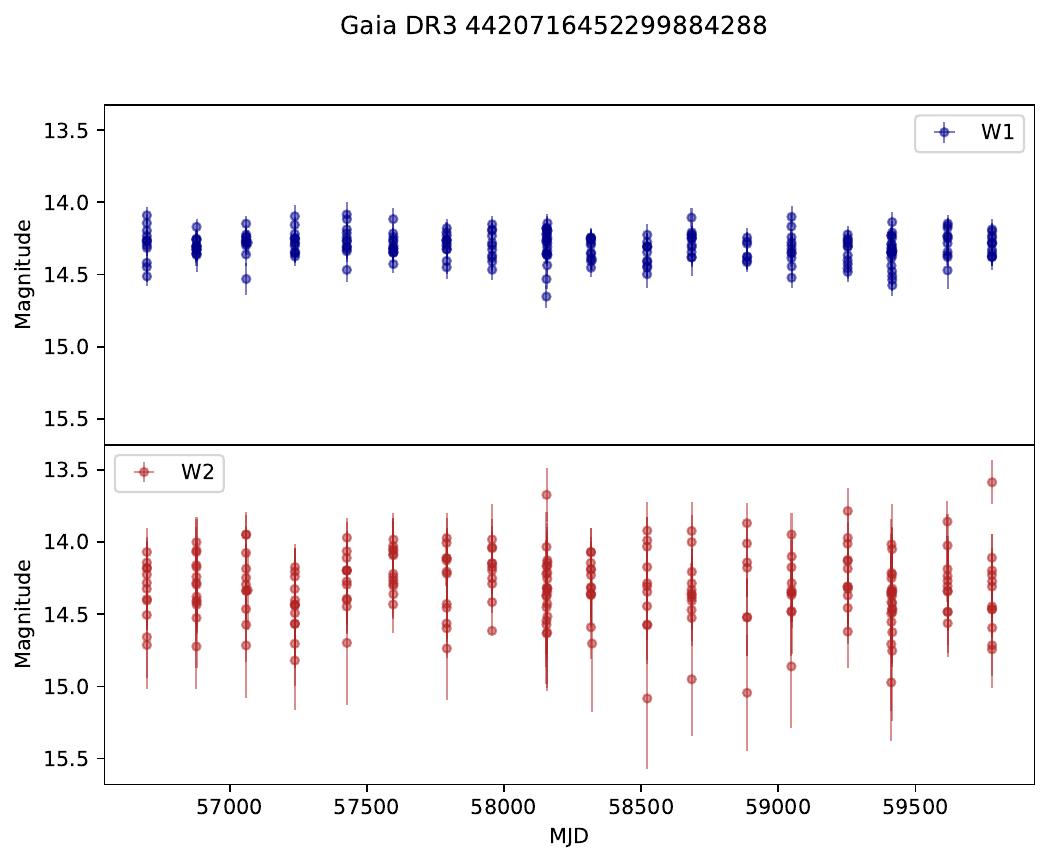}
    \caption{\textsl{NeoWISE} time-series for two of our candidates, in $W1$ (blue) and $W2$ (red) bands. Left panel shows a candidate with a high percentile score with respect to the control sample. Right panel shows a candidate with a low percentile score.}
    \label{fig:neowise_examples}
\end{figure*}

We compare the Stetson index obtained by our candidates against a random control sample and a set of neighbors (computed with NN-color), shown on the left panel of Figure \ref{fig:stetson}. We observe a small overdensity for higher Stetson indexes in our candidates' set, highlighting these candidates as potentially (more) variable objects. The object with the highest Stetson index also corresponds to an object flagged as a variable from \allwise{}'s data \textit{var\_flag}. However, the rest of our candidates have low IR variability scores according to this flag. To further get a sense of the variability of our candidates, we also measure their Stetson index percentile score (right panel Figure \ref{fig:stetson}) with respect to the Stetson indexes of the control sample (in blue), and with respect to their 30 neighbors (dashed black line). 
\\Figure \ref{fig:neowise_examples} shows the \textsl{NeoWISE} time-series in both $W1$ and $W2$ for two of our candidates, one with a high percentile score (left, ``more variable'') and one with a low percentile score (right).

We did not identify any rising pattern as the one in \cite{kenworthy2023planetscollision} in our candidates' \textsl{NeoWISE} time-series. Those types of objects might not get detected by our current pipeline, depending on how \allwise{} and \unwise{} treat the aggregation of \textsl{NeoWISE} data, and when the rise would occur: we might get under-estimated magnitudes from either catalogs, or pre-rise magnitudes (from \allwise{}). Additionally, we removed candidates with a disagreement between the two catalogs: as \unwise{} used a longer observation span from \textsl{NeoWISE}, a strong change in an object magnitude could lead to disagreement with \allwise{} as well. A dedicated search for these specific "rising IR emissions" would therefore be relevant, using either \textsl{NeoWISE} or \textsl{unTimely} \citep{meisner2023untimely}.


On the optical side, we check the photometric variable flag provided by \gaia{} (computed in the $G$ band). 10 out of our 53 candidates are flagged as variable, the remaining have not been processed (flag not available, hence they can not be considered not-variable either). This is a significantly larger occurrence of variable objects ($\sim 19\%$) compared to the occurrence rate in our control sample ($\sim 7\%$), the ``neighbors'' sample ($\sim 2.5\%$), and our entire sample ($\sim 5.3\%$). Since variability is an indicator of potential youngness, this could hint that our sample tends to have younger stars than our control set. However, dimming and transit patterns have been observed in the optical for Sun-like stars with similar MIR-excess with ages of 300Myr \citep{kenworthy2023planetscollision} and 600Myr \citep{melis2021structureddisktransit}. A closer investigation of the variability and periodicity of our candidates would thus be necessary.
\\According to the variable classification pipeline \citep{rimoldini2023variableclassifgaia} provided by \gaia{}, one of our candidates is an eclipsing binary (\gaia{} DR3 6705842630930258816). Two objects are (tentatively) classified as RS Canum Venaticorum variable (\gaia{} DR3 3183786280737192704 and 2256421065353454464), one as a $\delta$ Scuti or $\gamma$ Doradus or SX Phoenicis star. Four remaining are flagged as ``solar-like''.

Using the cross-match of \gaia{} DR3 with known variable sources in the literature, provided by \cite{gavras2023variablecrossmatchDR3}, we identify two candidates classified as Eclipsing, one that was previously identified as such above by the \gaia{} variable flag and \gaia{} DR3 5260430146906778624.

Longer light curves in the optical can be obtained for some of these objects (30) from \textsl{TESS} mission. However, a deeper analysis of e.g. periodicity and conducting gyrochronology will require a careful pre-processing and detrending of those data, which we defer to future work. Observations from the Zwicky Transient Facility (ZTF) could also be leveraged (17 candidates) through the SNAD viewer \citep{malanchev2023snad} which also provides cross-matches across multiple catalogs and additional variability detection from various pipelines.


\subsection{Characterization of the Excesses through SED-fitting}

We now further investigate our curated set of 53 candidates to derive properties of the possible source of excess.

First, we use \ariadne{} package \citep{2022Ariadne} to perform Spectral Energy Distribution (SED)-fitting on each of our candidates. Given RA,DEC positions and optionally a \gaia{} DR3 identifier, \ariadne{} automatically retrieves the object's photometric observations from \gaia{} DR2, \allwise{}, \textsl{APASS}, \textsl{Pan-STARRS1}, \textsl{SDSS}, \tmass{}, \textsl{ASCC}, \textsl{IRAC/GLIMPSE}, \textsl{TESS}, \textsl{SKYMAPPER}, \textsl{GALEX}, and \textsl{Tycho-2}. \ariadne{} also incorporates priors for interstellar extinction parameters through dust maps. We use the default setup relying on SFD dustmaps \citep{1998SFD}.
The goal of \ariadne{} library is originally to improve SED-fitting through bayesian model averaging of different models. However, in the following of our analysis, we will use a single SED-model, the Castelli-Kurucz \citep{Castelli2004}. We also use the `fitzpatrick' extinction-law in \ariadne{} setup.
\ariadne{} then returns the best-fitted SED as well as estimated stellar parameters (Teff, logg, distance, $A_V$, metallicity and radius). \ariadne{} uses by default all observed photometry up to $W2$ for the fit, but not $W3$ and $W4$.

We follow this pipeline, using \gaia{} DR3 values as priors for \textsl{Teff} and \textsl{logg}. However, we force the removal of the \wise{} values. The motivation is two-fold: (i) we put the SED-modeling in a similar context as our model, (ii) since we will use the fitted SED to compute properties of the potential source of the excess, we do not want to bias the predicted values in the \wise{} bands.
\\
We obtain, for each of our candidates, SED-predicted values for all the \wise{} bands. Note that these values use potentially more observations at different wavelengths than our model, and rely on stellar model, extinction and other priors (contrary to our entirely data-driven predictions).

\begin{figure} 
    \centering
    \includegraphics[width=.95\linewidth]{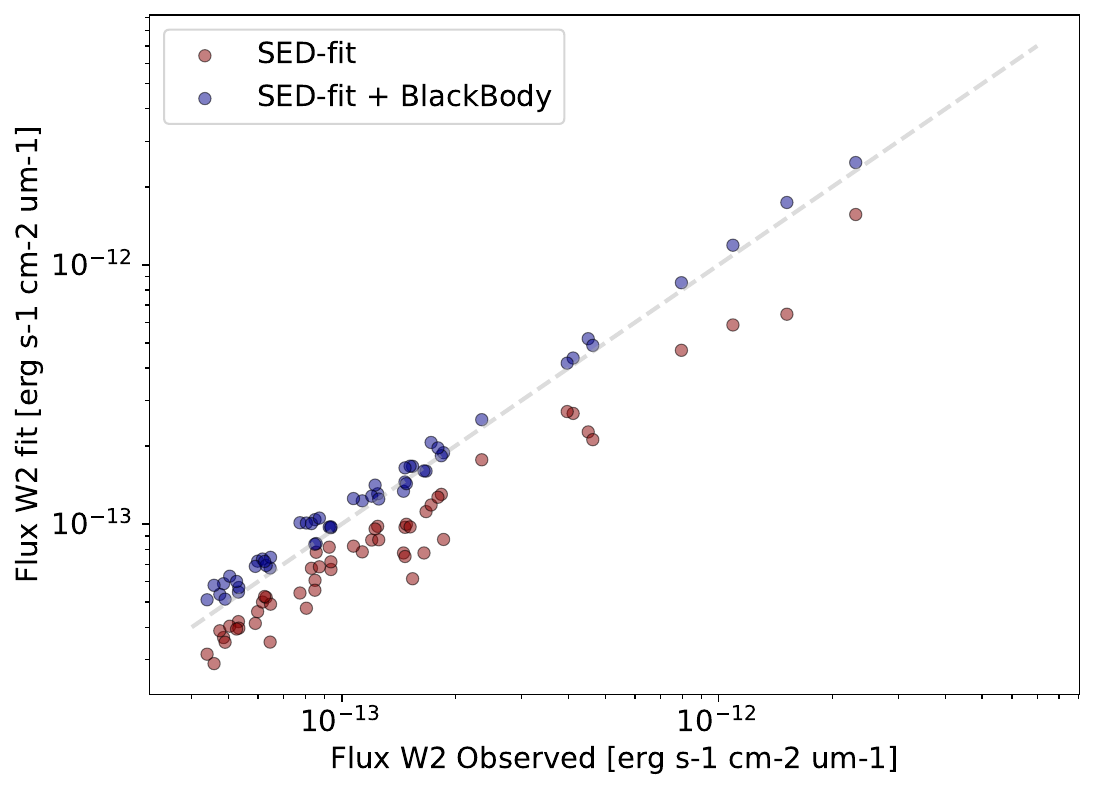}
    \caption{Observed flux (in $W2$ band) \textit{vs} predicted flux from \ariadne{}-SED only (red points) and \ariadne{}-SED $+$ a BlackBody (blue points), fitted on the flux residuals in bands $K_s, W1, W2$ and $W3$. The 1-to-1 line is depicted as a dashed grey line.}
    \label{fig:blackcorrection}
\end{figure}

\begin{figure*}[tp]
\centering
  \includegraphics[width=.8\linewidth]{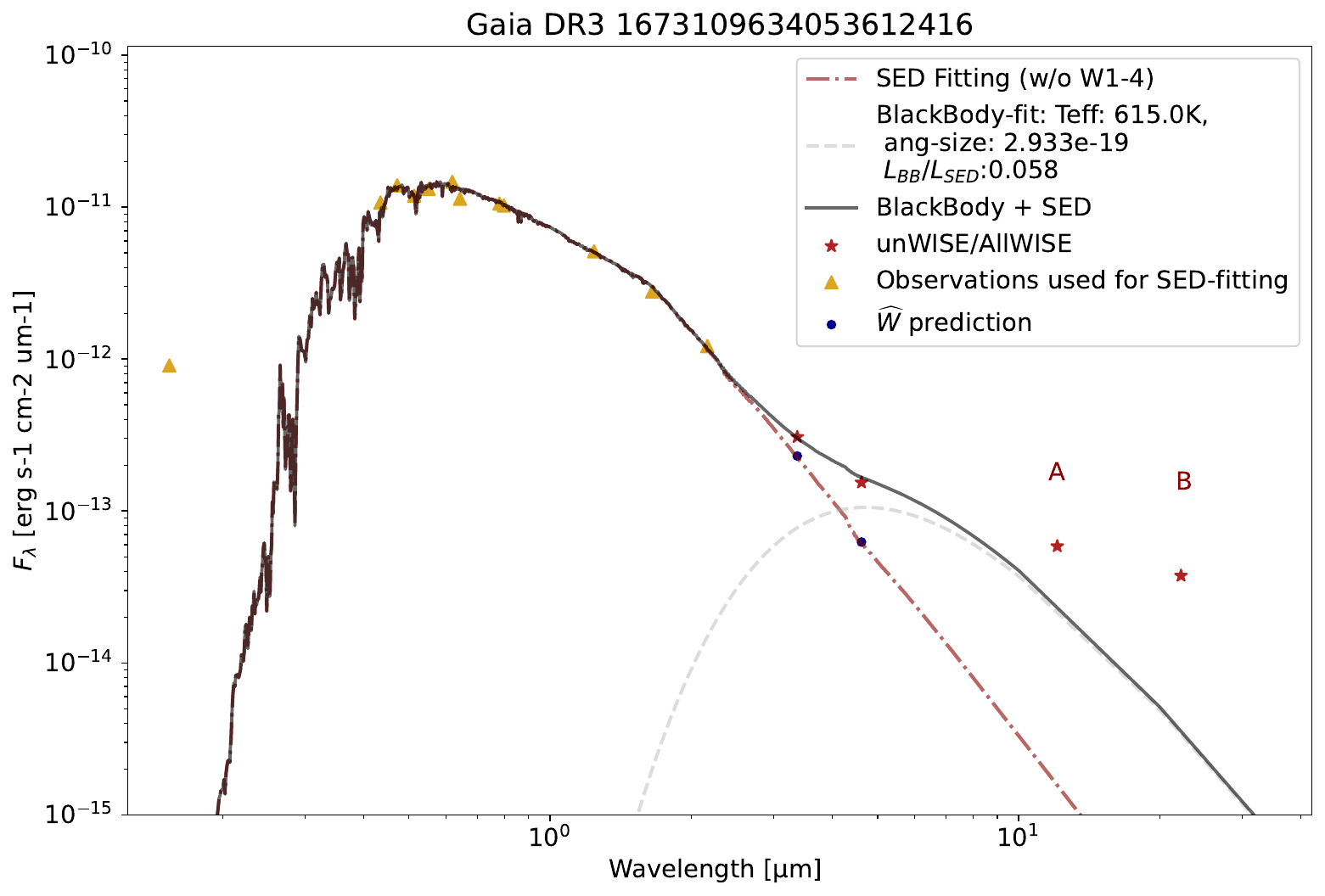}
  \caption{SED and BlackBody fitting for our excess candidates. Yellow triangles are the observations gathered and used by \ariadne{} to compute the SED (dashed dark red line). Red points are the \unwise{} (for $W1$ and $W2$) and \allwise{} (for $W3$ and $W4$) values. Blue points are the prediction of our model (averaged over the 7-folds). The best-fit blackbody is depicted as a grey dashed line. The final SED computed as \ariadne{} SED + BlackBody is depicted as solid black line. The letters above $W3$ and $W4$ are \allwise{} photometric quality flag.}
    \label{fig:SEDone}
\end{figure*}

Using the SED-predicted fluxes and the observed ones, we can estimate the properties of a blackbody fitted on the residuals. To do so, we build a grid of blackbody spectra with a range of temperatures from $100$K to $1500$K with a step of $5$K, and a range of angular sizes from $10^{-23}$ to $10^{-16}$ steradians with $200$ bins spaced evenly on log-scale\footnote{The luminosity of a blackbody being set by its surface area and temperature, the apparent flux is set by a solid angle here in steradians.}. For each candidate, we select the blackbody that minimizes the sum of the squared errors as: $argmin_{BB} \sum_{j}(F_{BB}(\lambda_j) - (F_{obs}(\lambda_j) - F_{SED}(\lambda_j)))^2 $ where $j$ denotes an index on wavelengths. We fit the blackbody using four wavelengths corresponding to $K_s, W1, W2$, and $W3$ bands. Because we optimize the sum of the squared errors in the flux units, the brighter bands are weighted more than the fainter ones in the fit. While we do not use $W3$ for our detection pipeline because it is less reliable, we choose to use it for the blackbody fitting to extend our wavelengths baseline. But it is important to note that some of our candidates have a low SNR flag in $W3$, which might lead to incorrect blackbody estimates. 
\\We show in Figure \ref{fig:blackcorrection} the flux value observed in $W2$ band versus the flux predicted by the SED-fit as red points, confirming excess for all of our candidates (lying under the 1-to-1 dashed line). The blue dots show the flux in $W2$ of the SED with the additional blackbody, against the observed flux. Our fitting procedure tends to slightly over-correct (some points lie above the 1-to-1 dashed line) in the $W2$ band, but not in $W1$ and $W3$.
\\Figure \ref{fig:SEDone} shows the SED from \ariadne{} (dashed red line), the black body fitted (dashed light grey line), and the combination of the two (black line), for one of our candidates.

\begin{figure*}
    \centering
    \includegraphics[width=.8\textwidth]{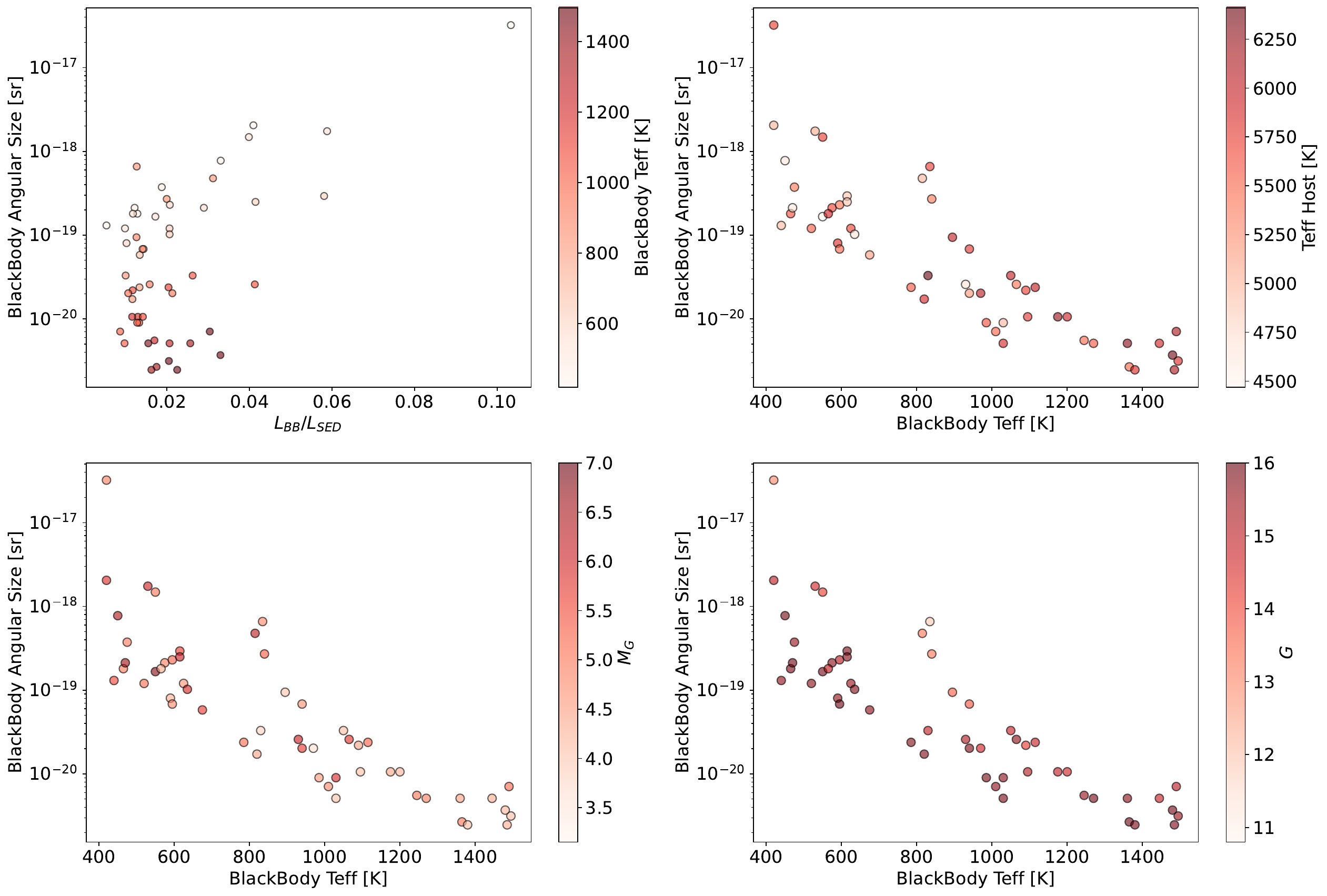}
\caption{Black bodies (BB) and stellar properties: Top left panel shows the fractional luminosity against the angular size, colored by the BB's temperature. Top right panel shows the BB's temperature against the angular size, colored by the effective temperature of the star provided by \gaia{}. Bottom panels show the same space colored by the absolute magnitude $M_G$ (left) and $G$ (right).}
\label{fig:BB_Hosts_ppty}
\end{figure*}

Figure \ref{fig:BB_Hosts_ppty} shows the different black bodies (BB) properties we obtain (bearing in mind that those are approximate fits) with our candidates' sample. The fractional luminosities $f_{BB}=\frac{L_{BB}}{L_*}$ (computed as the ratio of the integrated flux of the sole black body and the integrated flux of the predicted SED) range between $0.005$ and $0.1$, with an average of $0.02$. These fractional luminosities align with what has been observed and expected for Extreme Debris Disks candidates.

The SED and black-body fits for all of our candidates are shown in Figures \ref{fig:SED0}-\ref{fig:SED3}, along with their XP spectra (H$\alpha$ line indicated in blue) and imaging visualization of the objects in $W1$ and $W2$ bands using \allwise{}, and in the optical using the Digitized Sky Survey (DSS), on $3'$ square. We observe in these Figures a diversity of ``excess shape'' and strength: we see candidates with some extreme ``bumps'' in $W3$ -- which are often not well fitted by our black body estimates --, similar to EDDs candidates. However, as we also display the photometric quality flag of $W3$ and $W4$ from \allwise{} \footnote{{The quality flag is $A$ if a source is detected with a flux SNR $> 10$, $B$ if  $3 < \text{SNR} < 10$, C if $2 < \text{SNR} < 3$, and $U$ if $ \text{SNR} <2 $, indicating upper limit on magnitude: the magnitude of the given band is a 95\% confidence upper limit. All our excess candidates have an $A$ quality flag for $W1$ and $W2$.}}, we can see that for some of those objects, the $W3$ band might not be reliable. This might also impact the relevance of our BB parameters fit. We also want to point out that a blackbody, or a single blackbody, might not be the best way to represent the source of the excess. Some works use two blackbodies e.g. \cite{chen2014twoBBmodels}. We defer the investigation of more complex models to future work.

These candidates might interestingly fall in a region of debris disks evolution and parameter space that was scarcely populated previously and provide an interesting continuum of disks' evolution: \cite{wyatt2015evoldisks} proposed a five steps classification in the evolution of disks (from protoplanetary to debris), based on a number of a factor including flux ratio at different wavelengths (see Fig. 1 in \citealt{wyatt2015evoldisks}). Our candidates would fall around HD 141569 for the flux ratio at $12 \mu m$. However, the lack of reliable observations at higher wavelengths, dust mass estimates, and, as mentioned previously, reliable age estimates, prevent us from placing our candidates in this classification scheme accurately. Additionally, the work conducted in \cite{wyatt2015evoldisks}  focused on A stars, while our candidates are FGK stars, which might make direct comparison difficult. This motivates nonetheless further observations on this sample to better connect and understand those potential objects as potentially ``transitional'' or undergoing extreme and rare events.

\subsection{Recovery Rate of Injections}

\begin{figure*}
    \centering
    \includegraphics[width=.7\linewidth]{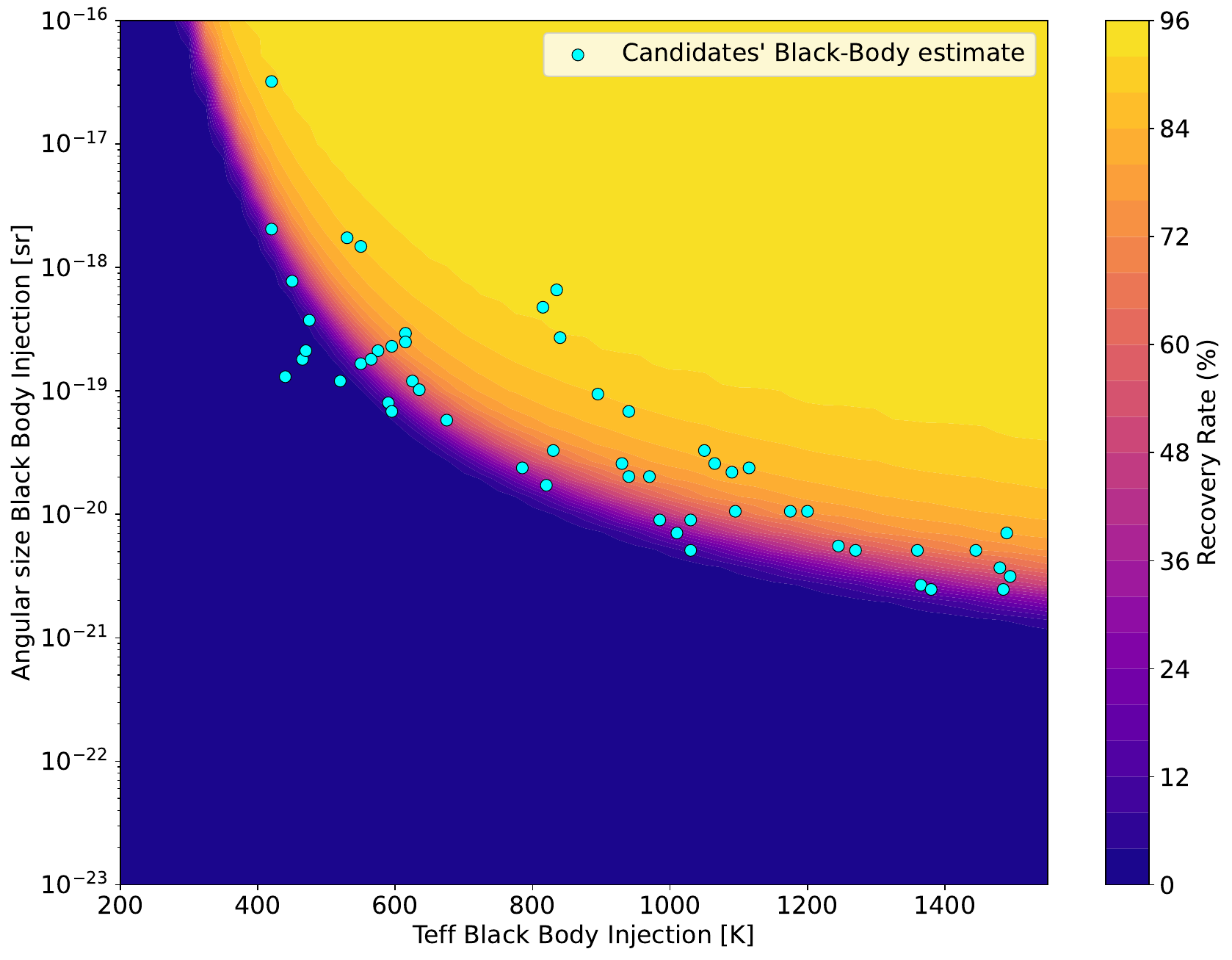}
    \caption{Map of the recovery rate of Black Body injections in a sample of stars similar to our candidates ($\sim$1,500 stars). Our candidates' BB estimates are shown as cyan dots.}
    \label{fig:recoveryinj}
\end{figure*}

As a sanity check, we perform a set of injections to evaluate the recovery rate of our pipeline depending on BB's parameters. We select as a sample of stars the set of neighbors (in color space using NN-color) of our candidate, $\sim 1500$ objects. We create a grid of BB parameters, where the temperature ranges from $200$ to $1575K$ by step of $25K$, and the angular size ranges from $10^{-23}$ to $10^{-16}$ with 50 bins in log-scale. Each star in the sample is injected with all BBs, creating an artificially injected dataset of 4.3M objects. We only consider here the recovery as passing the cut of the first three masks used in this paper relying on prediction error (Eq \ref{eq:mean_error_cut}), the confidence of error compared to similar objects (Eq \ref{eq:mean_errorMAD_cut}) and the precision criterion (Eq \ref{eq:foldmad_cut}). Figure \ref{fig:recoveryinj} shows the resulting recovery rate for each temperature-angular size bin, and our candidates as the cyan points. 
Some of our candidates' estimated BB parameters lie in a low or no-recovery region: this is very likely explained by the fact that our BB's parameters estimation is done with four observations, and often a rather approximated fit (as can be seen in Figures \ref{fig:SED0}-\ref{fig:SED3}).  

It is also important to note that our selection pipeline relies in the first cut on thresholds (Equations \ref{eq:mean_error_cut}-\ref{eq:foldmad_cut}): relaxing these choices and threshold might lead to more detections and higher recovery rates of fainter MIR-excess, but also potentially to an increase in the false detection rate. 

\subsection{Occurence rate}
\label{sec:occurencerate}

Our injection analysis combined with our candidates does not however directly allow to make statements on the occurrence rate of MIR-excess events in FGK main sequence stars, due, among other things, to the threshold steps of our pipeline and our data-driven approach. However, analyzing the rate of detection of infrared \textit{deficit} can help interpret our detection rate in the excess direction. As we assume that MIR deficits are not physically possible, investigating if and when we reach the detection of MIR-deficit gives us an idea of when we reach the ``noise'' level of our model, and the possible false-positive or contamination rate.

With our threshold values, looking for infrared deficits of similar strength (that is, reversing the cuts and values) in the opposite direction as done with  Eq. \ref{eq:mean_error_cut} and \ref{eq:mean_errorMAD_cut}, leads to a preliminary set of 13 ``deficit'' candidates (compared to 127 in the excess direction). This could be interpreted, to some extent, as a possible false-positive rate in our preliminary excess sample of $\sim 10\%$. 
\\When we apply the additional cuts of Table \ref{tab:cuts}, all the 13 candidates are eliminated. This, besides confirming the relevance of these checks to remove false detection, also tends to indicate a relatively high purity in our excess final sample. 
\\Conversely, replicating the pipeline using upper percentiles to compute the thresholds for infrared-deficit detection, leads to a final set of 7 candidates (compared to our 53 candidates), 3 of which can easily be discarded from imaging as being contaminated, the others being extremely small deviation (see Appendix \ref{sec:appendix:deficit} for the Figures). 

This combined with the injection analysis seems to indicate that our percentile approach for our threshold choice allows good coverage for recovery while not leading to too much false-positive contamination. Our (uncorrected)  ``occurrence'' rate (53 candidates out of $\sim$ 5 million stars, or $\sim 0.001 \%$) is much lower than the ones suggested in e.g. \cite{kennedy2013exozodi} (1 in 10,000 for old ($>$Gyr) systems --e.g. like BD+20 307--, and $~1\%$ for younger ($<$120Myr) systems). However, their estimates are for IR excess at 12 $\mu$m, while our detection pipeline uses observations at $\sim 2-3 \mu$m. 


\section{Discussion}
\label{sec:discu}

We proposed and executed a data-driven approach for the discovery of infrared (IR) excesses around stellar objects, more specifically among main-sequence FGK stars. We used robust machine learning methods to bypass stellar modeling to estimate the ``expected'' mid-IR (MIR) magnitudes, and we designed a set of curated statistics derived from the predictions and the predictions' quality to define a set of criterion cuts. We combined those with additional cuts to prevent as much as possible false detection and artificial excess due, e.g., to contamination. We obtained a set of 53 candidates that display interesting (M)IR excess. A few of those objects appear to be young stars (showing high $H\alpha$ emission). A significant fraction of our candidates seem to have variability in the optical, and some in the mid-infrared. This can also indicate youth, but a proper age estimation using gyrochronology and a variability analysis is required. Using SED-fitting and fitting black bodies on the flux residuals in 4 bands ($K_s, W1, W2, W3$), we compute black body properties and obtain estimated fractional luminosities coherent with the ones associated with Extreme Debris Disks candidates and planetary collision candidates. These objects form an interesting set of candidates to investigate further, as they show a variety in their excess' morphologies, some of which might require more refined models than a simple single blackbody.

It is natural to follow this work by gathering follow-up observations of the candidates, to confirm their nature (or discard potential confounding factors), to obtain better estimates of the properties of those disks, their compositions and their underlying processes. For example, \cite{moor2024} recently investigated the presence of solid-state features around 10 $\mu m$ in eight EDDs, and divided them into silica or silicate-dominated groups. Another key aspect will be to perform a deeper analysis to obtain reliable age estimates: the most interesting objects (as in: the ones that would challenge current models the most or would be the least expected) would be non-young. Additionally, identifying potential trends (if any) between those excesses and stellar ages might help better understand and constrain the underlying physical processes leading to such excesses. 

As we mentioned in the introduction, technosignatures such as Dyson spheres or swarms would lead to infrared excesses, of potentially different morphologies depending on the level of completion of the sphere/swarm. \cite{suazo2024} recently conducted a search where they looked for specific types of IR-excess matching the Dyson spheres' models they defined. It would be interesting to follow a similar analysis on our set of candidates --which was found through a more flexible search, and thus all might not necessarily match the IR-excess morphologies of DS-- to investigate if some of our candidates are potential matches and at which level of completion. 

Another natural extension of this work is to investigate other stellar types. These kinds of IR-excess have also been observed around A-stars (e.g. in \cite{melis2013Astars}), or in M-dwarfs (e.g. in the techno signature search of \cite{suazo2024}). Replicating this work for different stellar types separately could lead to interesting observations, in terms of event rates and, further down the line, disk types.

While such a pipeline is relevant to identify a set of outliers candidates, it can only find (contextual) outliers according to \textit{the data}. Making population statements (e.g. ``true'' occurrence rate of planetary collisions) is not trivially doable with this approach, as it relies heavily on a series of threshold cuts to build the candidate sample. Additionally, since our model computing the ``expected (M)IR magnitude'' is data-driven, we cannot directly use the error in prediction as the true IR excess (or deficit) for the full sample to investigate for instance the fraction of different levels of IR excess. We also note that, in principle, the prediction errors can be due mainly to variance in the data (and hence the model not being perfectly accurate) and might not necessarily indicate ``interesting anomalousness''. Here we can benefit from the fact that we are looking for the largest deviations, we expect (and have) a small sample of outliers to investigate, and we have other methods to validate an excess after detection. We also benefit from the fact that in this application, symmetric opposite anomalies (in the deficit side) are not physically possible. Thus we can use those as a probe for false-positive detection and/or reaching ``noise'' level with our model. This would not always be available in other applications or general cases.

\foreach \index in {0, ..., 2} 
{
\begin{figure*}
    \centering
    \includegraphics[width=\linewidth]{Figures/SED_img_spectra_\index.pdf}
    \caption{SED and BlackBody fitting for our excess candidates. Yellow triangles are the observations gathered and used by \ariadne{} to compute the SED (dashed dark red line). Red points are the \unwise{} (for $W1$ and $W2$) and \allwise{} (for $W3$ and $W4$) values. Blue points are the prediction of our model. The best-fit blackbody is depicted as grey dashed line. The final SED computed as \ariadne{} SED + BlackBody is depicted as a solid black line. Each plot also displays imaging from \allwise{} ($W1,W2$) and \textsl{DSS} of a $3'$ square centred on the source candidate. 
    The bottom part of each plot is the \textsl{XP} absolute spectra from \gaia{} DR3 computed using \textsl{Gaiaxpy} of the object (in red). The spectras of the 30 nearest-neighbors (using NN-mag) are in grey. H$\alpha$ line is shown in blue. The letters above $W3$ and $W4$ are \allwise{} photometric quality flag. 
    }
    \label{fig:SED\index}
\end{figure*}
}

\begin{figure*}
    \centering
    \includegraphics[width=.6\linewidth]{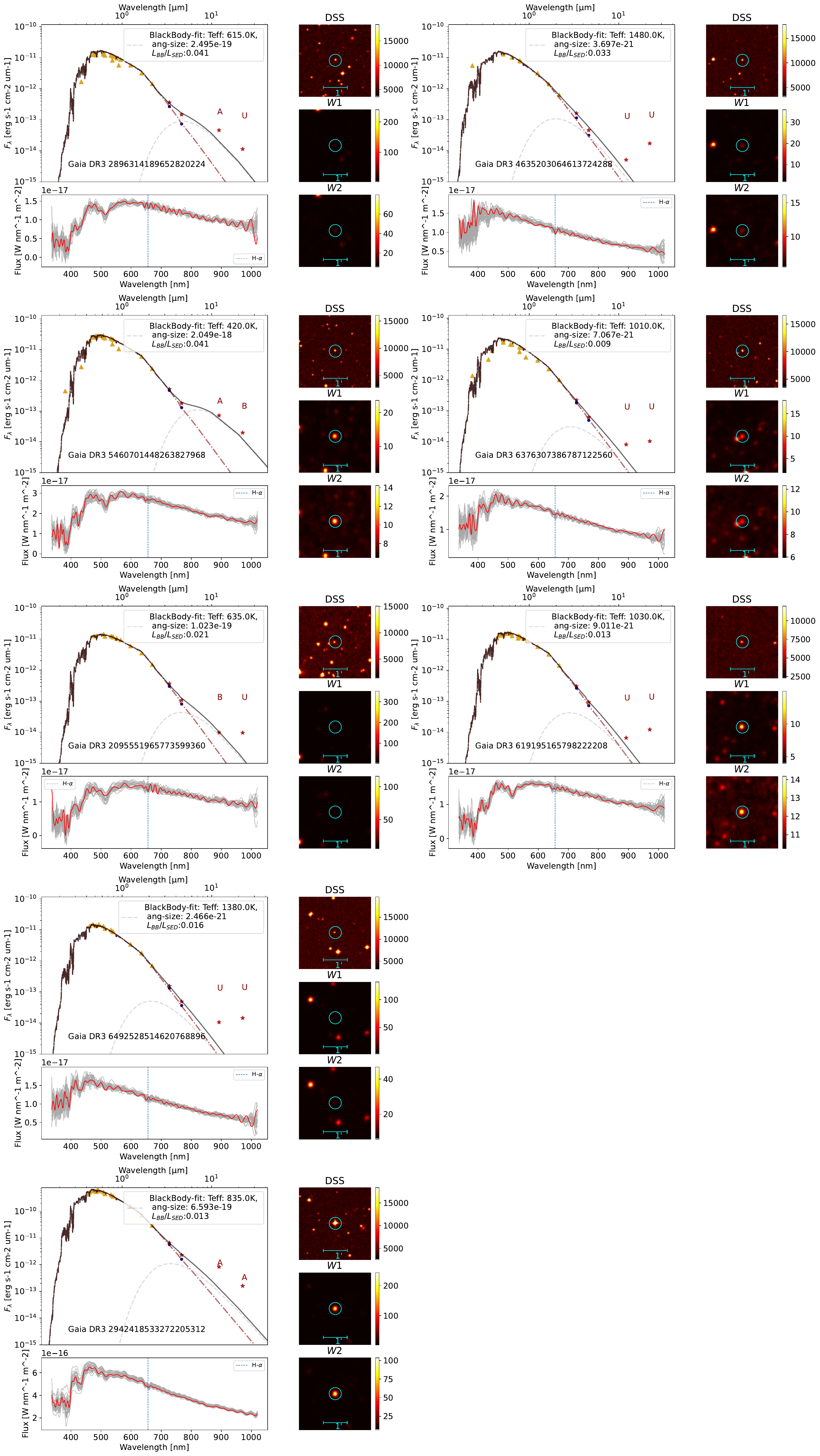}
    \caption{SED and BlackBody fitting for our excess candidates. Yellow triangles are the observations gathered and used by \ariadne{} to compute the SED (dashed dark red line). Red points are the \unwise{} (for $W1$ and $W2$) and \allwise{} (for $W3$ and $W4$) values. Blue points are the prediction of our model. The best-fit blackbody is depicted as a grey dashed line. The final SED computed as \ariadne{} SED + BlackBody is depicted as a solid black line. Each plot also displays imaging from \allwise{} $W1,W2$ and \textsl{DSS} of a $3'$ square centered on the source candidate. 
    The bottom part of each plot is the \textsl{XP} absolute spectra from \gaia{} DR3 computed using \textsl{Gaiaxpy} of the object (in red). The spectra of the 30 nearest neighbors (using NN-mag) are in grey. H$\alpha$ line is shown in blue. The letters above $W3$ and $W4$ are \allwise{} photometric quality flag.}
    \label{fig:SED3}
\end{figure*}

\section*{Acknowledgments}
It is a pleasure to thank Jason Wright,
Phil Armitage, and Erwan Allys for valuable discussions, as well as our anonymous reviewer and Carl Melis for their useful suggestions.

This work has made use of data from the European Space Agency (ESA) mission {\it Gaia} (\url{https://www.cosmos.esa.int/gaia}), processed by the {\it Gaia} Data Processing and Analysis Consortium (DPAC,
\url{https://www.cosmos.esa.int/web/gaia/dpac/consortium}). Funding for the DPAC has been provided by national institutions, in particular the institutions participating in the {\it Gaia} Multilateral Agreement.

This publication makes use of data products from the Two Micron All Sky Survey (doi: 10.26131/IRSA2 \citep{tmassDOI}), which is a joint project of the University of Massachusetts and the Infrared Processing and Analysis Center/California Institute of Technology, funded by the National Aeronautics and Space Administration and the National Science Foundation.

This publication makes use of data products from the Wide-field Infrared Survey Explorer, which is a joint project of the University of California, Los Angeles, and the Jet Propulsion Laboratory/California Institute of Technology, and NEOWISE, which is a project of the Jet Propulsion Laboratory/California Institute of Technology. WISE and NEOWISE are funded by the National Aeronautics and Space Administration.

This publication also makes use of \unwise{} data product (doi:10.26131/IRSA525 \citep{unwiseDOI, 2019unwise}) and \allwise{} data products (doi:  10.26131/IRSA1 \citep{allwiseDOI}).

This research uses services or data provided by the Astro Data Lab at NSF's National Optical-Infrared Astronomy Research Laboratory. NOIRLab is operated by the Association of Universities for Research in Astronomy (AURA), Inc. under a cooperative agreement with the National Science Foundation.

The Flatiron Institute is a division of the Simons Foundation.



\software{Astropy \citep{astropy:2013, astropy:2018, astropy:2022}, numpy \citep{numpy}, ARIADNE \citep{2022Ariadne}, scikit-learn \citep{sklearn}, astroquery \citep{astroquery}, matplotlib \citep{matplotlib}, pyvo \citep{pyvo}, scipy \citep{scipy}, pandas \citep{pandas}, dustmaps \citep{2018dustmap}, gaiaXPy \citep{gaiaxpy}.}

\newpage
\appendix

\section{Candidates Table}
\label{sec:appendix:canditable}

Table \ref{tab:candisummary} provides a summary view of our candidates with \gaia{} DR3 source ID, magnitudes in $G$ and $W2$ bands, variable flags from \allwise{} and \gaia{} (when available, \textit{NaN} does not mean a candidate is not variable), Stetson index $S_J$, effective temperatures of our blackbody fit, fractional luminosities according to blackbody fit, pseudo-equivalent width of the H$\alpha$ line from \gaia{} (negative value corresponds to emission), \gaia{}'s age estimate, and the distance of the closest radio source within 15" if any.


\startlongtable
\begin{deluxetable*}{ccccccccccc}
\tablecaption{Summary table of Candidates \label{tab:candisummary} }
\tablehead{\colhead{Gaia DR3 ID} & \colhead{$G$} & \colhead{$W2$} & \colhead{var-flag} & \colhead{var-flag} & \colhead{$S_J$} & \colhead{BB Teff} & \colhead{$L_{IR}/L*$} & \colhead{EWH$\alpha$} & \colhead{Gaia Age } & \colhead{Radio source}\\ \colhead{} & \colhead{(mag)} & \colhead{(mag)} & \colhead{AllWise} & \colhead{Gaia} & \colhead{} & \colhead{(K)} & \colhead{} & \colhead{} & \colhead{(Gyr)} & \colhead{dist. (arcsec)}}
\startdata
5085509597958589184 & 15.76 & 14.02 & 00nn & NaN & 0.18 & 785 & 0.01 & 0.12 & 1.00 & NaN \\
3462639162734302464 & 15.52 & 13.74 & 11nn & NaN & 0.30 & 625 & 0.02 & 0.12 & 8.43 & NaN \\
6502798709139275136 & 15.61 & 13.99 & 00nn & NaN & 0.26 & 590 & 0.01 & 0.17 & 6.88 & NaN \\
6836154657401562112 & 14.24 & 12.54 & 110n & NaN & 0.25 & 1090 & 0.01 & 0.06 & 9.45 & NaN \\
2886776063720502144 & 13.69 & 11.93 & 000n & NaN & 1.07 & 895 & 0.01 & 0.14 & 5.26 & NaN \\
6705842630930258816 & 15.76 & 13.64 & 11nn & VARIABLE & 1.21 & 940 & 0.02 & -0.03 & 3.20 & NaN \\
387557473467906432 & 14.92 & 13.33 & 11nn & NaN & 0.30 & 830 & 0.01 & 0.18 & 1.33 & 0.51 \\
3889059420143391744 & 15.87 & 13.23 & 11nn & NaN & 0.08 & 550 & 0.02 & 0.06 & 9.79 & NaN \\
2144442781092733952 & 15.68 & 13.70 & 000n & NaN & 0.31 & 575 & 0.03 & -0.03 & 1.71 & NaN \\
1406113335695903872 & 15.63 & 13.67 & 11nn & NaN & 0.08 & 675 & 0.01 & 0.08 & 4.13 & NaN \\
1815157499757887488 & 15.81 & 14.25 & 00nn & NaN & 0.10 & 820 & 0.01 & 0.10 & 2.25 & NaN \\
1704765570250298496 & 15.85 & 14.15 & 00nn & NaN & 0.09 & 465 & 0.01 & 0.01 & 5.55 & NaN \\
4710409831751731712 & 15.68 & 13.94 & 11nn & NaN & 0.35 & 1360 & 0.03 & 0.23 & NaN & NaN \\
3772493492333504768 & 15.54 & 13.54 & 00nn & NaN & 0.16 & 475 & 0.02 & 0.22 & 10.98 & NaN \\
1245496161713883904 & 15.65 & 13.97 & 11nn & NaN & 0.21 & 520 & 0.01 & 0.07 & 1.37 & NaN \\
3017430205815280384 & 14.21 & 11.83 & 011n & VARIABLE & 1.00 & 550 & 0.04 & 0.09 & 4.31 & NaN \\
302802517288053632 & 15.96 & 13.55 & 11nn & NaN & 0.16 & 470 & 0.01 & 0.02 & 4.35 & NaN \\
3221650540619783040 & 14.96 & 13.24 & nnnn & NaN & 0.23 & 1175 & 0.01 & 0.15 & 0.30 & NaN \\
4420716452299884288 & 15.84 & 14.27 & 00nn & NaN & 0.01 & 1030 & 0.01 & 0.23 & 6.62 & NaN \\
6830573291565067648 & 15.88 & 14.21 & 00nn & NaN & 0.08 & 985 & 0.01 & 0.05 & 10.09 & NaN \\
907249585730646016 & 14.69 & 12.91 & 011n & NaN & 0.32 & 1200 & 0.01 & 0.24 & 5.71 & 0.30 \\
1800491354668025344 & 15.96 & 14.15 & 00nn & NaN & 0.09 & 1365 & 0.02 & 0.10 & 10.33 & NaN \\
1813144053445574784 & 15.10 & 13.05 & 110n & NaN & 0.32 & 1490 & 0.03 & 0.15 & NaN & 1.80 \\
1673109634053612416 & 15.85 & 12.99 & 110n & NaN & 0.23 & 615 & 0.06 & -0.02 & NaN & 10.65 \\
6578449850774068224 & 13.30 & 11.21 & 000n & NaN & 1.29 & 840 & 0.02 & 0.10 & 5.21 & NaN \\
5260430146906778624 & 14.80 & 12.87 & 99nn & VARIABLE & 2.64 & 1115 & 0.02 & -0.74 & NaN & NaN \\
709086185604311552 & 15.97 & 14.16 & 11nn & NaN & 0.02 & 595 & 0.01 & 0.29 & 11.91 & NaN \\
6809873615078147456 & 14.83 & 12.79 & 00nn & NaN & 0.11 & 1050 & 0.03 & 0.16 & 4.25 & NaN \\
6359523376149191936 & 14.64 & 11.80 & 011n & NaN & 0.25 & 530 & 0.06 & -0.00 & 11.46 & NaN \\
6770961795500483200 & 15.30 & 12.80 & 100n & NaN & 0.30 & 930 & 0.02 & -0.04 & NaN & NaN \\
6405864801961715968 & 15.85 & 14.36 & 00nn & NaN & 0.08 & 1485 & 0.02 & 0.15 & 0.64 & NaN \\
352632796577124992 & 15.00 & 13.27 & nnnn & NaN & 0.29 & 565 & 0.01 & 0.14 & 11.85 & NaN \\
3004530769758177280 & 13.00 & 10.51 & 1100 & NaN & 1.56 & 420 & 0.10 & -0.76 & 5.79 & NaN \\
6877999557398569600 & 15.66 & 12.92 & 00nn & NaN & 0.11 & 1065 & 0.04 & 0.14 & 2.12 & NaN \\
2916199883936404864 & 14.91 & 13.22 & 110n & NaN & 0.35 & 1445 & 0.02 & 0.12 & 4.81 & NaN \\
344897216879901312 & 14.71 & 13.05 & 110n & VARIABLE & 0.23 & 970 & 0.01 & 0.11 & 4.84 & 0.95 \\
6727198140355736448 & 13.82 & 11.97 & 110n & NaN & 0.66 & 940 & 0.01 & 0.12 & 6.69 & NaN \\
1052596673505827840 & 15.87 & 14.04 & 00nn & NaN & 0.03 & 1270 & 0.02 & 0.16 & 6.43 & 0.25 \\
3183786280737192704 & 13.31 & 10.87 & 420n & VARIABLE & 1.11 & 815 & 0.03 & 0.06 & NaN & NaN \\
2256421065353454464 & 15.65 & 13.63 & 42nn & VARIABLE & 0.56 & 440 & 0.01 & -0.01 & NaN & NaN \\
1938237827799536512 & 15.89 & 13.04 & 000n & NaN & 0.30 & 450 & 0.03 & 0.11 & NaN & NaN \\
5456960634826855168 & 15.56 & 13.64 & 001n & VARIABLE & 0.19 & 1245 & 0.02 & 0.07 & 9.51 & NaN \\
2770417668730232192 & 15.14 & 13.01 & 11nn & NaN & 0.13 & 595 & 0.02 & 0.05 & 12.13 & NaN \\
4941459109269026560 & 15.54 & 13.94 & 00nn & NaN & 0.10 & 1495 & 0.02 & 0.05 & 7.44 & NaN \\
6612104218072332160 & 15.17 & 13.61 & 11nn & NaN & 0.14 & 1095 & 0.01 & -0.03 & 9.90 & NaN \\
2896314189652820224 & 15.79 & 13.04 & 000n & VARIABLE & 0.74 & 615 & 0.04 & -0.01 & 3.47 & NaN \\
5460701448263827968 & 15.05 & 12.83 & 011n & VARIABLE & 0.81 & 420 & 0.04 & 0.03 & 8.99 & NaN \\
2095551965773599360 & 15.78 & 13.39 & 11nn & NaN & 0.31 & 635 & 0.02 & 0.03 & NaN & NaN \\
6492528514620768896 & 15.88 & 14.24 & 00nn & NaN & 0.05 & 1380 & 0.02 & 0.04 & 7.47 & NaN \\
2942418533272205312 & 11.86 & 10.06 & 0111 & VARIABLE & 1.58 & 835 & 0.01 & 0.10 & 4.69 & NaN \\
4635203064613724288 & 15.97 & 14.31 & 01nn & NaN & 0.11 & 1480 & 0.03 & 0.34 & 0.35 & NaN \\
6376307386787122560 & 15.64 & 13.98 & 10nn & NaN & 0.06 & 1010 & 0.01 & 0.06 & 13.19 & NaN \\
619195165798222208 & 15.71 & 13.54 & 01nn & NaN & 0.22 & 1030 & 0.01 & 0.02 & 3.91 & NaN \\
\enddata
\end{deluxetable*}

\section{ADQL Query for Gaia DR3 data selection and cross-match}
\label{sec:appendix:adql}

Gaia ADQL query for data selection and cross-match with \allwise~ and \tmass~ of FGK stars. This query has to be looped over ranges of \textit{start\_index} and \textit{end\_index} using \gaia{} DR3 \textit{random\_index} provided, to cover the entire \gaia{} data.

\begin{verbatim}
SELECT dr3.*, xmatch.*, allw.*, tmassmatch.*, xjoin.*, tmass.*, 
dr3_all.duplicated_source 
FROM gaiadr3.gaia_source_lite AS dr3 
JOIN gaiadr3.gaia_source AS dr3_all USING (source_id)
JOIN gaiadr3.allwise_best_neighbour AS xmatch USING (source_id) 
JOIN gaiadr1.allwise_original_valid AS allw
ON xmatch.allwise_oid = allw.allwise_oid 
JOIN gaiadr3.tmass_psc_xsc_best_neighbour AS tmassmatch USING (source_id) 
JOIN gaiadr3.tmass_psc_xsc_join AS xjoin USING (clean_tmass_psc_xsc_oid)
JOIN gaiadr1.tmass_original_valid AS tmass 
ON xjoin.original_psc_source_id = tmass.designation 
WHERE (dr3.random_index BETWEEN start_index AND end_index 
AND dr3.parallax_over_error > 10.0 
AND dr3.teff_gspphot BETWEEN 4000 AND 7000 
AND dr3.phot_g_mean_mag < 16 AND dr3.ruwe < 1.4 
AND dr3.ebpminrp_gspphot IS NOT NULL AND allw.cc_flags = '0000' 
AND tmass.ks_msigcom < .2 
AND tmass.h_msigcom < .2 AND tmass.j_msigcom < .2 AND tmass.ph_qual = 'AAA' 
AND tmassmatch.number_of_neighbours = 1  AND  xmatch.number_of_mates = 0 
AND xmatch.number_of_neighbours = 1 AND tmassmatch.number_of_mates = 0 
AND xmatch.angular_distance < .15 AND tmassmatch.angular_distance < .15)

\end{verbatim}

\section{Infrared Deficit}
\label{sec:appendix:deficit}
As noted in Section \ref{sec:occurencerate}, using the same values used for Eq. \ref{eq:mean_error_cut} and \ref{eq:mean_errorMAD_cut} in a symmetric way to look for infrared \textit{deficit}, i.e. looking for deficit of similar ``strength'', leads to only 13 deficit candidates in the preliminary cut (as opposed to 127 in the excess direction). All of them then get discarded either by the secondary cuts (Table \ref{tab:cuts}).
\\
Thus, we have no IR-deficit candidates that are symmetrically similar in strength to our IR-excess candidates.

\begin{figure*}
    \centering
    \includegraphics[width=.6\linewidth]{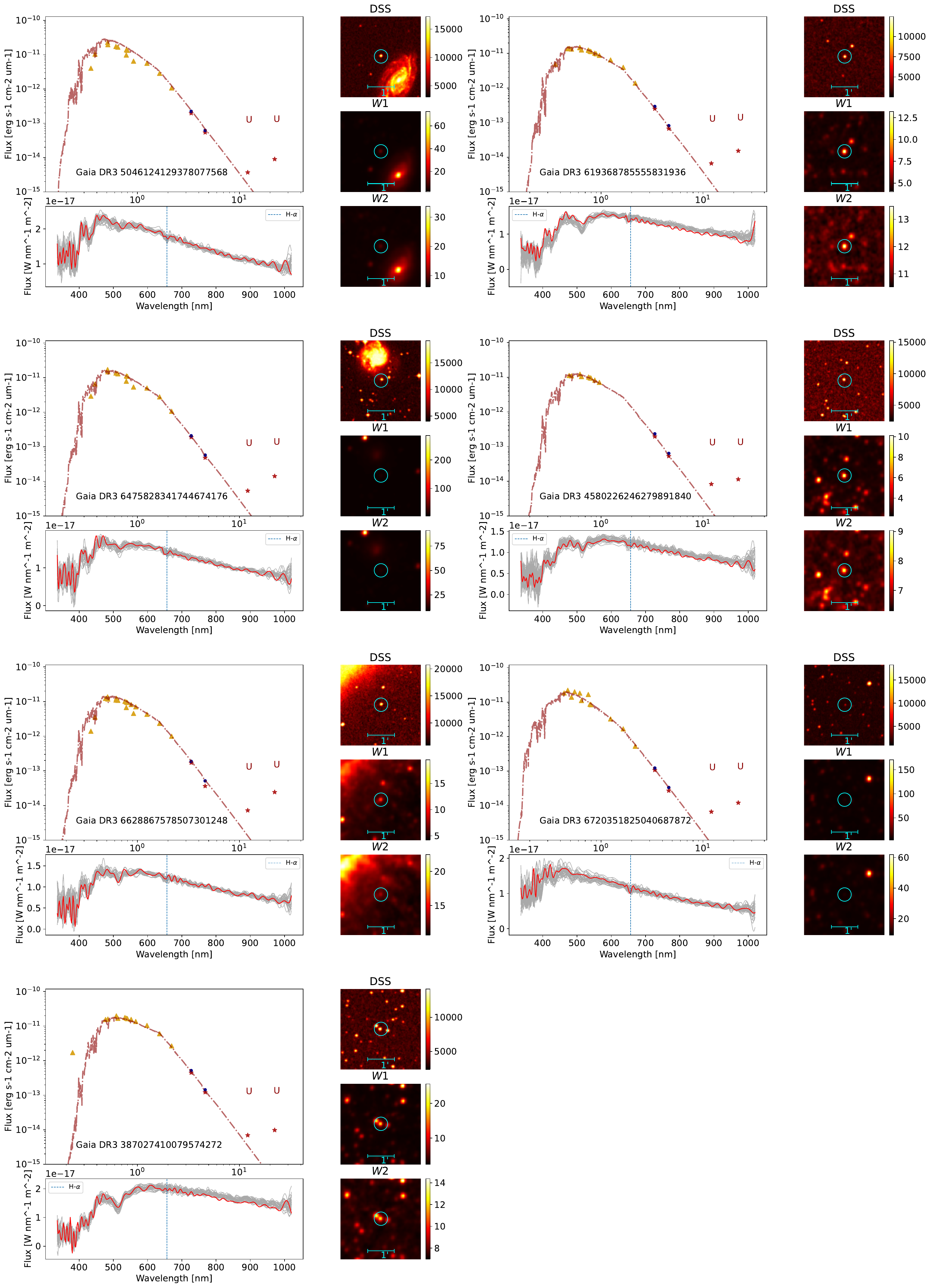}
    \caption{SED, imaging and XP spectra of \textbf{IR-deficit candidates}. }
    \label{fig:SED-deficit}
\end{figure*}

Applying a similar methodology where Eqs \ref{eq:mean_error_cut} and \ref{eq:mean_errorMAD_cut} using upper percentiles, i.e. looking at the ``most deficit'' objects (but not necessarily as strong in deficit as our excess candidates), we obtain 83 preliminary candidates (\textit{vs} 127 for excess). Applying all the secondary cuts leaves only 7 candidates (shown Figure \ref{fig:SED-deficit}). This highlights the efficiency of those secondary cuts to potentially remove false candidates.
\\Three of those have likely contamination from their environment shown by the imaging. They appear as a relatively small deficit in $W1$ and $W2$ according to the SED-fit too. For all of these examples, the $W3$ and $W4$ land above the SED-fit. However, they also all have a quality flag $U$ (meaning that the source measurement had an SNR $< 2$, and the profile-fit magnitude is a 95\% confidence upper limit).

\section{Overlap with previously detected IR-excess}
\label{sec:appendix:overlap}
We check our pipeline selection and list of candidates against existing IR-excess candidates from the literature. We summarize this search in Table \ref{tab:overlap}. For each reference investigated, we retrieve catalogs from either SIMBAD or Vizier, retrieve the \gaia{} DR3 match from SIMBAD if there is one (second column), and check if those objects are in our full 4.9M sample (third column). We indicate which of those objects make it through our preliminary cut (combining Equations \ref{eq:mean_error_cut},\ref{eq:mean_errorMAD_cut} and \ref{eq:foldmad_cut}) (fourth column) and through the complete pipeline (fifth column). Note that some objects in those catalogs might not be matched yet with a DR3 identifier and could thus actually appear in our sample. 

The excesses most similar to our candidates are likely to be Extreme Debris Disks candidates, and we indeed find that these objects are the ones that most often make it through our preliminary cut. Out of 23 objects matched from the catalog available on SIMBAD\footnote{We note that the original paper however mentions 17 candidates.}, 5 appear in our full sample, and they all make it through our preliminary cut, confirming the ability of our pipeline to detect such IR-excesses. Two of those are then discarded by the crowding cut, one is discarded by the proper-motion disagreement cut, and one by the k-NN ``density'' cut. One remains in our final set of candidates, \gaia{} DR3 2942418533272205312 (bottom left plot in Figure \ref{fig:SED3}).
\\
Out of the few other IR-excess candidates appearing in our 4.9M sample (42 in total), only one passes through the preliminary cut (from the Reserved catalog by \cite{cotten2016census}), but is discarded by the crowding criterion.

This encourages us to believe that we explored a large sample of stars relatively disjointed from previous searches, both in the population sample analyzed (little overlap in the main catalog) and in the type of excess detected. It is relevant to note that the IR-excesses in the literature were detected in $W3$ and $W4$ bands, and hence might not produce an excess in the MIR detectable by our pipeline. It also requires stringent cuts on the quality and SNR of $W4$, which eliminates a large number of objects. Our detection pipeline allows for the discovery of extreme outliers (similar to potential EDDs) by relying on $W1$ and $W2$ bands only, while being rather conservative in further removing potential false detection.

\begin{deluxetable*}{cccccc}[tp]
\tablecaption{Table of overlap between existing IR-excess catalogs and our sample and set of candidates. We retrieve the DR3 identifiers counterparts of the candidates using SIMBAD and indicate the number of objects in our full sample, the number of objects passing the preliminary cut and the full pipeline selection.}
\tablehead{\colhead{Reference} & \colhead{Objects with} & \colhead{Objects in} & \colhead{Preliminary Cut} & \colhead{Full Cut} \\{} & {DR3 match} & {our 4.9M catalog} & {} & {} \label{tab:overlap}} 
\startdata
 Extreme Debris Disk \citep{moorEDD} & 23 & 5 & 5 & 1 \\
\cite{cotten2016census}, Prime & 491 & 0 & 0 & 0 \\
\cite{cotten2016census}, Reserved & 1218 & 18 & 1 & 0 \\
 Sun-Like stars \citep{cruz2014sunlike} & 199 & 1 & 0 & 0 \\
Main-Sequence \citep{dacosta2017wiseIR} & 210 & 6 & 0 & 0 \\
Tycho-\gaia{} DR1 \cite{McDonald2017TychoGaia} & 3836 & 16 & 0 & 0 \\
 Cold Debris Disk Spitzer \citep{ballering2013coldDD} & 225 & 0 & 0 & 0 \\
 T-tauri \citep{manzo2020protoTTauri} & 784 & 1 & 0 & 0 \\
 Herbig Ae/Be \citep{arun2019herbig} & 121 & 0 & 0 & 0 \\
 Cepheids \citep{schmidt2015cepheid} & 27 & 0 & 0 & 0 \\
\enddata
\end{deluxetable*}

\bibliography{sample631}{}
\bibliographystyle{aasjournal}

\end{document}